\documentclass[%
 aip,
pop,
amsmath,
amssymb,
reprint,%
]{revtex4-1}

\usepackage{tikz}
\usepackage{graphicx}
\usetikzlibrary{positioning}
\usepackage{dcolumn}
\usepackage{bm,todonotes}
\usepackage{pifont}

\begin{document}


\title[Evolution of the ECDI in 2D]{Evolution of the electron cyclotron drift instability in two-dimensions}
\author{Salomon Janhunen}
\email{salomon.janhunen@usask.ca.}
\affiliation{University of Saskatchewan, 116 Science Place, Saskatoon, SK S7N 5E2 Canada}
\author{Andrei Smolyakov}
\affiliation{University of Saskatchewan, 116 Science Place, Saskatoon, SK S7N 5E2 Canada}
\author{Dmytro Sydorenko}
\affiliation{University of Alberta, 3-235 Centennial Centre for Interdisciplinary Science Edmonton, AB T6G2E9, Canada}
\author{Marilyn Jimenez}
\affiliation{University of Saskatchewan, 116 Science Place, Saskatoon, SK S7N 5E2 Canada}
\author{Igor Kaganovich}
\affiliation{Princeton University / Princeton Plasma Physics Lab, 100 Stellarator Rd, Princeton 08543-0451, USA}
\author{Yevgeny Raitses}
\affiliation{Princeton University / Princeton Plasma Physics Lab, 100 Stellarator Rd, Princeton 08543-0451, USA}



\date{\today}

\begin{abstract}
The Electron Cyclotron Drift Instability (ECDI) driven by the electron $E\times B$ drift in partially magnetized plasmas is investigated with highly resolved particle-in-cell simulations. The emphasis is on two-dimensional effects involving the parallel dynamics along the magnetic field in a finite length plasma with dielectric walls.
It is found that the instability develops as a sequence of growing cyclotron harmonics demonstrating wave breaking and complex nonlinear interactions, being particularly pronounced in ion density fluctuations at short wavelengths. At the same time, nonlinear evolution of fluctuations of the ion and electron density, as well as the anomalous electron current, shows cascade toward long wavelengths. Tendency to generate long wavelength components is most clearly observed in the spectra of the electron density and the anomalous current fluctuations. An intense but slowly growing mode with a distinct eigen-mode structure along the magnetic field develops at a later nonlinear stage enhancing the tendency toward long wavelength condensation. The latter mode having a finite wavelength along the magnetic field is identified as the Modified Two-Stream Instability (MTSI). It is shown that the MTSI mode results in strong parallel heating of electrons.
\end{abstract}

\pacs{Valid PACS appear here}
\keywords{Suggested keywords}
\maketitle

%

%

\section{\label{sec:intro}Introduction}
Partially magnetized plasmas immersed in crossed $E\times B$ fields are used in various devices such as Hall thrusters for electric propulsion. Such plasmas are subject to a number of instabilities that affect device operation -- and in particular -- the level of anomalous transport that is typically found to be orders of magnitude larger than the classical (collisional) transport. The nature of the anomalous transport (mobility) is still poorly understood and has been attributed to several candidate instabilities that may interact with each other to bring about the observed levels  of anomalous transport. The electron cyclotron drift instability (ECDI) driven by the electron $E\times B$ drift, and independent of any plasma  gradients and collisions, has been recently actively discussed as a possible candidate \cite{Mikellides2016,LafleurPoP2016b,BoeufJAP2017}.  

In earlier works \cite{ForslundPRL1970,GaryJPP1970,LampePRL1971,StenzelPRL1973}, electron cyclotron instabilities have been studied in relation to turbulent plasma heating by the electric current perpendicular due to the relative electron-ion drift. Electron cyclotron instability driven by ion beams was also identified as a possible source of anomalous resistivity explaining the width of collisionless shock waves, in particular  in space conditions; for more recent work and references see Refs.  \onlinecite{MatsukiyoJGR2009,MuschiettiJGR2013,MuschiettiAdvSpR2006}. 

In the context of the anomalous transport in Hall thrusters, the cyclotron instability driven by the electron $E\times B$ drift was studied in 1D simulations \cite{DucrocqPoP2006,BoeufFP2014,BoeufJAP2017,LafleurPoP2016a},  2D axial-azimuthal simulations \cite{AdamPoP2004,BoeufIEPC2017,CochePoP2014}, and 2D radial-azimuthal simulations \cite{CroesPSST2017,HaraIEPC2017,HeronPoP2013}.
 Many of these works focused on the possibility that ECDI simply becomes the ion sound instability analogous to the case of unmagnetized  plasma\cite{CavalierPoP2013,LafleurPoP2016a}. We have shown in our previous nonlinear simulations that the transition to ion sound (which for the 1D case is only possible due to nonlinear diffusion in the short wavelength regime) does not occur\cite{JanhunenPOP2018} and the instability is driven by the dominant $m=1$ cyclotron resonance.  We also found in our previous 1D simulations that when larger azimuthal length is used, anomalous transport cascades to low-$k$ modes. Curiously in Ref.~\onlinecite{LafleurPoP2016a} the authors see the emergence of a large-scale structure in their simulation when they use a larger simulation box, but reject it as an artifact.

In this paper, using highly resolved particle-in-cell simulations, we study instabilities and transport in the 2D (azimuthal-radial) geometry. Periodic boundary conditions are used in the azimuthal direction, along the $E\times B$ drift. The magnetic field is in the radial direction bound by the dielectric wall boundaries. Curvature effects of the channel are not included in this work.

In 2D geometry a new class of unstable mode appears for finite values of the the wave number $k_z$ along the magnetic field, namely the Modified Two-Stream Instability (MTSI) \cite{McBridePF1972}. For larger values of $k_z$, the unstable mode looks similar to the unmagnetized ion sound \cite{GaryJPP1970a,GaryJPP1970,CavalierPoP2013}. In this paper, we study the linear and nonlinear evolution of the interacting ECDI and MTSI modes, their saturation and associated turbulent transport for typical conditions of  a Hall thruster. Our simulations demonstrate that like in the 1D case, the instability is driven by nonlinear cyclotron resonance modes that dominate the anomalous transport, and that the cascade to long wavelengths observed in 1D simulations is further enhanced by the linear long wavelength instabilities that occur when finite $k_z$ is allowed. Moderate values of the anomalous electron current (of the order of $\Omega =\Omega_{ce}/\nu_{eff}\simeq 200$ are obtained in nonlinear stage similar to the 1D case\cite{JanhunenPOP2018}. An important new result is strong parallel electron heating due to the modes with a finite $k_z$.

The paper is structured as follows: we discuss the 2D linear regime and the unstable normal modes therein, and show that they appear as expected in fully non-linear simulations in the very early part of the simulation. The early non-linear saturation processes are discussed, such as mode competition between the ECDI harmonics and apparent coupling to the modified two-stream instability (MTSI). In the development of the strong turbulence regime, we show how the MTSI compresses the ECDI wave packet and produces large fluxes to the sheath with accompanied rapid heating of the parallel temperature. We discuss the spectral cascade of the anomalous current, the features of the anomalous current as a function of the radial (parallel to magnetic field) direction, and the time evolution of the overall anomalous current. Finally, the sheath losses and decay of the plasma column in the absence of sources are discussed, and a summary follows. We discuss technical details such as numerical parameters and analysis methods in the Appendix.

\section{\label{sec:instability}Linear features of the Electron Cyclotron Drift-Instability}

The electron drift cyclotron instability (ECDI) occurs in partially magnetized $E\times B$ plasma due to the significant $E\times B$ flow of electrons with respect of ions. It is convenient to discuss the characteristics of the ECDI with reference to the linear dispersion relation. We consider the electrostatic waves with $\mathbf{v}_{0}=\mathbf{E\times B}/B^2$ streaming of electrons across a uniform magnetic field $\mathbf{B}$, with unmagnetized ions, in homogeneous unbounded plasma. The two-dimensional linear dispersion equation has the form \cite{GaryJPP1970a}
\begin{equation}
\epsilon \left( \omega ,\mathbf{k}\right) =1+\epsilon _{i}\left( \omega ,\mathbf{k}\right) +\epsilon _{e}\left( \omega ,\mathbf{k}\right) =0,
\label{disp}
\end{equation}
where $\epsilon_{e}$ and $\epsilon_{i}$ are the electron and ion susceptibilities
\begin{gather}
\epsilon _{i}=-\frac{1}{2k^{2}\lambda _{D_{i}}^{2}}\mathrm{Z}^{^{\prime }}\hspace{-0.25ex}\left(\frac{\omega}{\sqrt{2}kv_{i}}\right),\label{ions}\\
\begin{aligned}
\epsilon_{e}=\frac{1}{k^{2}\lambda _{De}^{2}}\left[ 1+\frac{\omega -\mathbf{k}\cdot\mathbf{v}_{0}}{\sqrt{2}k_{z}v_{e}}\sum\limits_{m=-\infty}^{\infty}e^{-b}\mathrm{I}_{m}\left( b\right)\right.\\\left.\mathrm{Z}\left( \frac{\omega -\mathbf{k}\cdot \mathbf{v}_{0}+m\Omega_{ce}}{\sqrt{2}k_{z}v_{e}}\right) \right],
\end{aligned}\label{el}
\end{gather}%
where $b=k_{y}^{2}\rho_{e}^{2}$, $\rho_{e}^{2}=v_{e}^{2}/\Omega _{ce}^{2},$
$v_{e,i}^{2}=T_{e,i}/m_{e,i}$, $\lambda_{De,i}^{2}=\epsilon_{0} T_{e,i} / n_{0} q_{e,i}^{2}$, $Z(\xi )$ is the plasma dispersion function, $I_{m}(x)$ is the modified Bessel function of the 1st kind, $\mathbf{B} = B_{0} \widehat{\mathbf{z}}$ is the magnetic field in the $z-$ direction,  $\mathbf{E}=E_{0}\widehat{ \mathbf{x}}$ is the external electric field in the x-direction, so that $\mathbf{v}=\mathbf{E\times B}/B^{2}=v_{0}\widehat{\mathbf{y}}$ is in the $y$ direction (or, against if $v_0=-E_0/B_0<0$). Then, $k_z$ and $k_y$ are the components of the wave vector $\mathbf{k}$ along the magnetic field and in the $\mathbf{E\times B/}$ directions, $k\equiv\left\vert \mathbf{k}\right\vert =\sqrt{k_{z}^{2}+k_{y}^{2}}$. Given the physical parameters, it remains to assign $k_z$ and $k_y$ values; solving the dispersion relation provides us with the frequency $\omega$, which generally gets a complex value.

For the geometry of the Hall thruster with a radial magnetic field, we define a few auxiliary quantities and relations that aid in the following discussions: $k_0=\Omega_{ce}/v_{0}$, $k_z=2\pi n_z/L_r$, $k_y=2\pi n_y/l_{\theta}$, where $L_r$ is the extent of the system in the radial (along the magnetic field) direction, $l_\theta$ is the extent of the system in azimuthal direction, so the quantum numbers $n_z$ and $n_y$ characterize the radial and azimuthal wave vectors that satisfy the boundary conditions. The local Cartesian coordinates $y,z$ correspond to the $\theta, r$ coordinates of the coaxial Hall thruster. It is important to emphasize however, that while the $y$-direction is periodic in our simulations, periodicity is not imposed in the radial direction. The eigenmode structure in radial direction is formed self-consistently by the mode parallel dynamics and by the sheath effects at $z=0$ and $z=L_r$. The ensuing mode structure will be discussed below. In this work we assume the gap to be a straight box, for simplicity.

In the limit of cold ions where $\omega>k v_{i}$, the ion response becomes
\begin{equation}
\epsilon_{i}=-\frac{\omega_{pi}^{2}}{\omega^{2}},
\label{ionc}
\end{equation}
where $\omega_{pi}^{2}=e^{2}n_{0}/\varepsilon_{0}m_{i}$ is the ion plasma frequency.

In one-dimensional case, $k_{z}\rightarrow 0$, $k=k_{y}$, the
dispersion equation (\ref{disp}) takes the form
\begin{equation}
\begin{aligned}
\epsilon_{e}&=\frac{1}{k^{2}\lambda _{De}^{2}}\left[ 1-\exp \left(
-k^{2}\rho _{e}^{2}\right) I_{0}\left( k^{2}\rho _{e}^{2}\right) \right.\\
&-2\left.\left( \omega -kv_{0}\right) ^{2}\sum\limits_{m=1}^{\infty }\frac{%
\exp \left( -k^{2}\rho _{e}^{2}\right) I_{m}\left( k^{2}\rho _{e}^{2}\right)
}{\left( \omega -kv_{0}\right) ^{2}-m^{2}\Omega _{ce}^{2}}\right].
\end{aligned}\label{eq:dispersion-magn}
\end{equation}
The form of Eq.~(\ref{eq:dispersion-magn}) emphasizes the role and the interaction of different cyclotron harmonics. Note that there is no resonance for the $m=0$ harmonic, while all higher harmonics with $m=1,2,..$ are resonant at $\left( \omega-kv_{0}\right) ^{2}=m^{2}\Omega_{ce}^{2}$. In the cold electrons limit $T_{e}\rightarrow 0$, only $m=0$ and $m=1$ harmonics contribute and the dispersion relation reduces to the Buneman magnetized plasma instability driven by transverse current \cite{Buneman1962}
\begin{equation}
1-\frac{\omega _{pi}^{2}}{\omega ^{2}}-\frac{\omega_{pe}^{2}}{(\omega
-kv_{0})^{2}-\Omega _{ce}^{2}}=0.  \label{bi}
\end{equation}
The instability is a result of reactive coupling of the electron (Doppler shifted) upper hybrid mode  $\left( \omega -kv_{0}\right) ^{2}=\omega_{pe}^2+ \Omega_{ce}^{2}$ with the short wavelength ion oscillations $\omega ^{2}=\omega_{pi}^{2}$. In the long wavelength low frequency limit, $\left( \omega,kv_{0}\right) < \Omega_{ce}$, equation (\ref{bi}) describes the lower-hybrid modes, $\omega _{LH}^{2}=\Omega _{ci}\Omega _{ce}. $ The contribution of higher $m>1$ harmonics (which are absent for $T_{e}=0$) grows with electron temperature and has the maximum at  shorter wavelengths $k^{2}\rho _{e}^{2}\simeq 1$ due to the  $\exp \left( -k^{2}\rho _{e}^{2}\right) I_{m}\left( k^{2}\rho _{e}^{2}\right)$ factors. The temperature effects also add dispersion to the lower hybrid modes $\omega ^{2}=\Omega _{LH}^{2}\left( 1+k^{2}\rho _{e}^{2}\right)$, which eventually becomes the high frequency ion sound for $k^{2}\rho _{e}^{2}\gg 1$, $\omega ^{2}=\Omega _{LH}^{2}k^{2}\rho _{e}^{2}\simeq k^{2}c_{s}^{2}$. In this limit, the contribution of the cyclotron harmonics decreases with $k \rho_{e}$:  $\exp \left( -k^{2}\rho _{e}^{2}\right) I_{m}\left( k^{2}\rho_{e}^{2}\right) \rightarrow 1/\left( k\rho _{e}\right)$, so that the real part of the electron susceptibility becomes 
\begin{equation}
\epsilon _{e}\simeq 1/\left( k^{2}\lambda _{De}^{2}\right),   \label{ee}
\end{equation}
and  equation (\ref{disp}) produces the ion sound mode 
\begin{equation}
\omega ^{2}=k^{2}c_{s}^{2}/\left( 1+k^{2}\lambda _{De}^{2}\right).
\label{is}
\end{equation}
The imaginary part in $\epsilon_{e}$ (neglected so far) originate in the series of cyclotron resonances. In the limit $k^{2}\rho_{e}^{2}\gg 1$, the infinite series of cyclotron resonances can be summed resulting to the imaginary contribution equivalent to the pole contribution $1/\left(\omega -kv_{0}\right)$ as for the case of unmagnetized electrons. Thus, even for strictly perpendicular propagation, in the $k^{2}\rho _{e}^{2}\gg 1$ limit, one has the ion sound instability as in the unmagnetized plasma case. This case is directly related to the resolution of the Landau-Bernstein paradox: the sequence of Bernstein modes which are undamped in magnetized plasma result in collisionless Landau damping when $B\rightarrow 0$.

Another kind of instability occurs near the resonances $\left( \omega -kv_{0}\right) ^{2}\simeq m^{2}\Omega_{ce}^{2}$. This is a strong (fluid) reactive instability due to coupling of the ion and electron modes \citep{LashmorePhysA1970}, facilitated by the Doppler shift. For cold electrons, only the $m=1$ exists, resulting to the Buneman instability described by Eq.~(\ref{bi}). For finite $T_{e}$, all higher modes with $m>1$ are present. In our previous work, it was shown that in 1D case a set of modes with  higher $m$ are excited, but eventually, via the linear (due to electron heating) and nonlinear effects, a dominant $m=1$ strongly coherent cnoidal wave appears.\cite{JanhunenPOP2018} The cyclotron resonance nature of the mode, defined by the condition $\omega <k_{y}v_{0}\simeq \Omega_{ce}$, extends far into the nonlinear stage. Similar result were also obtained in other simulations relevant to space plasma conditions \cite{MatsukiyoJGR2009}. 

In 2D where plasma motion along the magnetic field is present, new regimes become possible due to finite values of the $k_{z}$. There have been a number of studies of the full linear dispersion equation with (\ref{ions}) and (\ref{el}), for example see Refs.~\onlinecite{ArefevTechPhys1970,GaryJPP1970,GaryJPP1970a,CavalierPoP2013}. One of the results of these studies is that for sufficiently large values of the parallel wave vector $k_{z}$, the solution of the dispersion relation (\ref{disp}) produces a mode which is close to the ion sound instability in unmagnetized plasma driven by the electron beam with $v_0$ velocity.

\begin{figure}[htp]
\includegraphics[width=\columnwidth,clip]{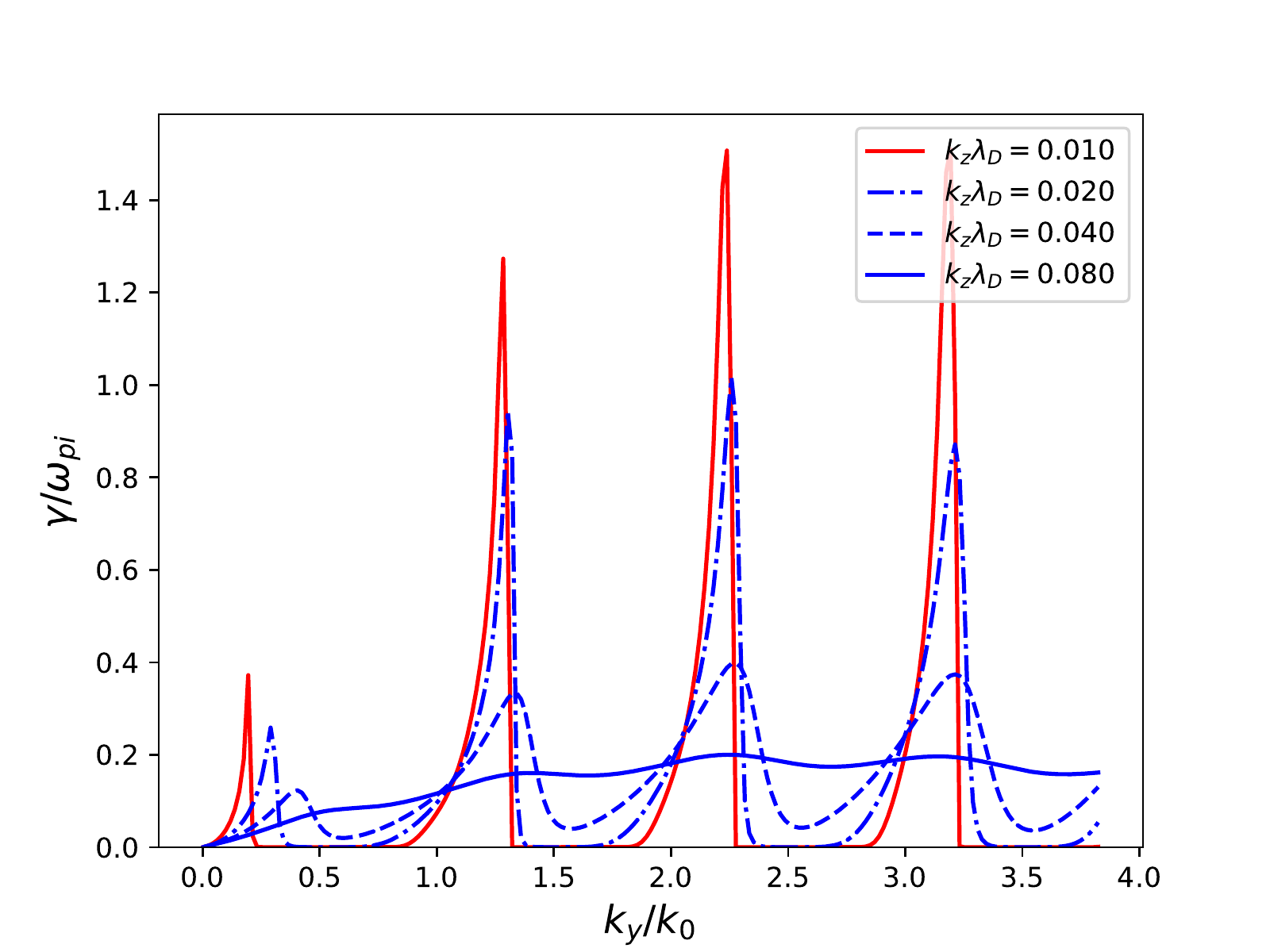}
\caption{Growth rate of instabilities for $k_z \protect\lambda_{De}$ of 0.01, 0.02, 0.04 and 0.08, respectively, found from the full dispersion equation (1)-(3). The first root from left is the MTSI (m=0), and subsequent roots are from $m=1,2,3,...$ ECDI resonances. Parameters for this figure are: $v_0=-369\,c_s$, $v_{Te}=489\,c_s$, $\Omega_{ce}=96.5\,\omega_{pi}$, $k_0\lambda_{De}=0.262$.}
\label{fig:gamma-kz}
\end{figure}

\begin{figure}[htp]
\includegraphics[width=\columnwidth,clip]{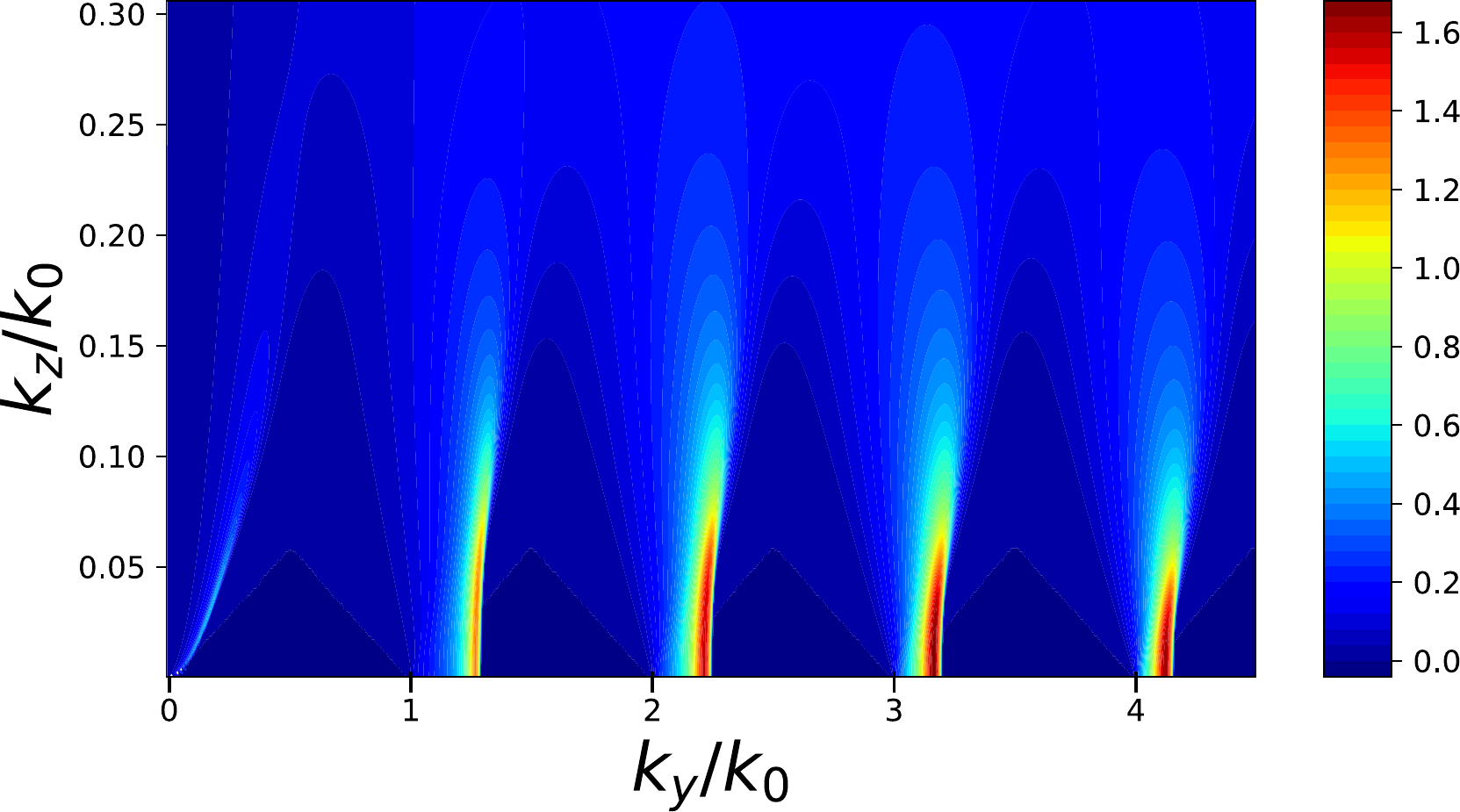}
\caption{Phase space plot of growth rate $\gamma(k_z,k_y)/\omega_{pi}$ for the full dispersion relation. Notice the modified two-stream instability (MTSI) in the sub-cyclotron low-$k$ region, indicating possible instability for simulations which are able to accommodate the mode in the azimuthal direction. Parameters for this figure are: $v_0=-369\,c_s$, $v_{Te}=489\,c_s$, $\Omega_{ce}=96.5\,\omega_{pi}$, $k_0\lambda_{De}=0.262$.}\label{fig:full-dispersion}
\end{figure}

\begin{figure}[htp]
\includegraphics[width=\columnwidth,clip]{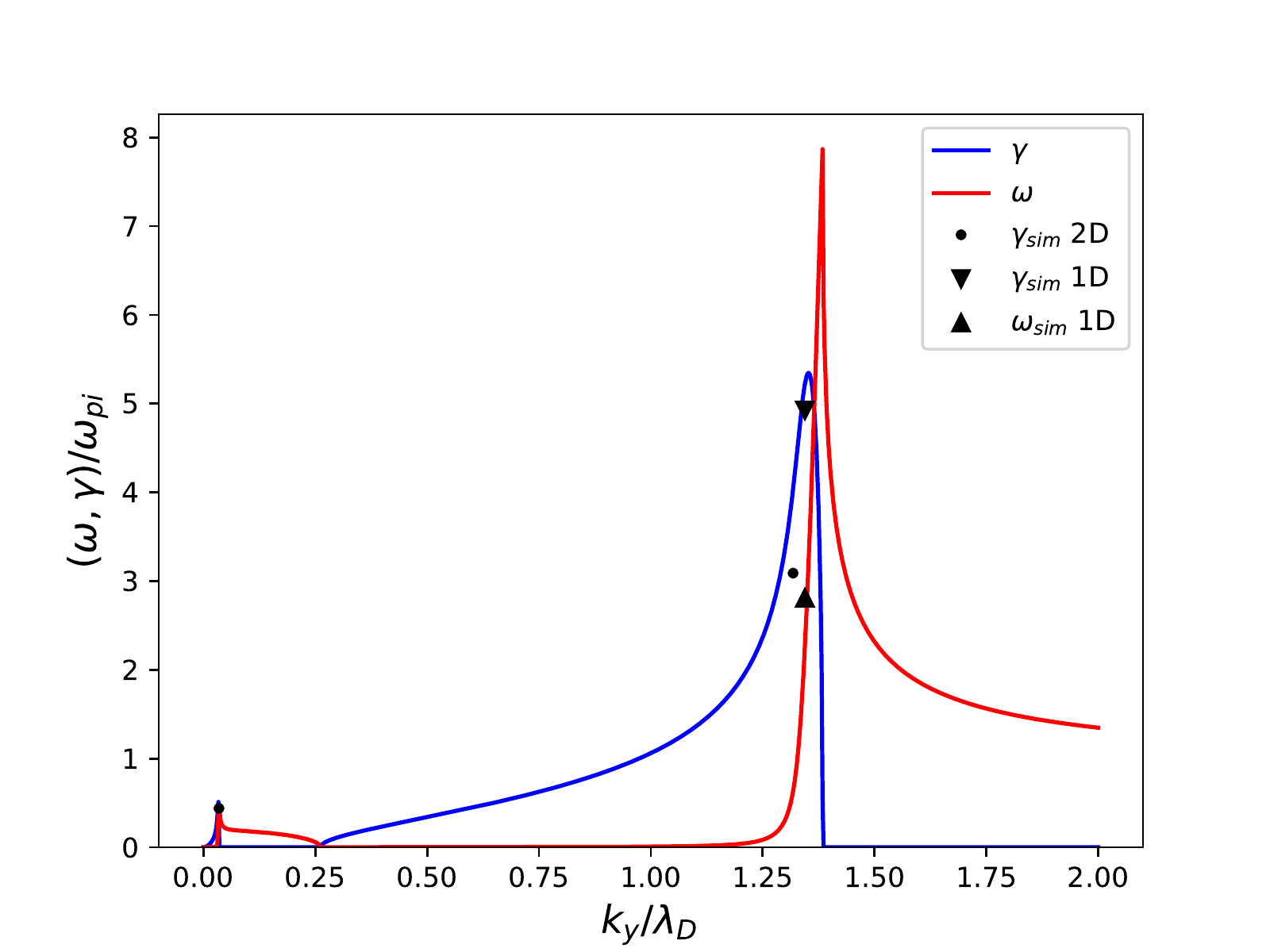}
\caption{Comparison of the growth rate from a 2D simulation and a 1D simulation with 40000 particles/cell, using $T_e=0.001$,  with the solution of the linear dispersion in the $T_e\rightarrow 0$ limit. Parameters for this figure are: $v_0=-36.9\cdot10^3\,c_s$, $v_{Te}=489\,c_s$, $\Omega_{ce}=96.5\,\omega_{pi}$, $k_z\lambda_{De}=0.00434$.}\label{fig:sim-gamma}
\end{figure}

The general behavior of the  growth rate of the instability is shown for several values of the $k_z$ parameter  in Fig.~\ref{fig:gamma-kz} as a function of the azimuthal wave vector $k_y$. 
We solve the dispersion relation numerically in Python\cite{Python} using the technique described in Ref.~\onlinecite{CavalierPoP2013}, where the solution is obtained through fixed point iteration using the relative error as a stopping condition. We use the convergence condition that $|1-\omega_{i+1}/\omega_{i}|<10^{-6}$ for $i\geq 15$.
The SciPy\cite{SciPy} Faddeeva function is used to get good numerical accuracy of the plasma dispersion function for a wide range of arguments. We show the first four roots obtained in the $(k_z,k_y)$ phase plane in figure~\ref{fig:full-dispersion} for a $10\,\text{eV}$ Xenon plasma with $n_e=10^{17}\,\text{m}^{-3}$, $B_0=0.02\,\text{T}$, and $E_0=20\,\text{kV/m}$, which are typical Hall thruster parameters. The normalization scheme for the dispersion relation solver code is the same as defined in Ref.~\onlinecite{CavalierPoP2013}, where frequencies are normalized to $\omega_{pi}$, velocities to $c_s$ and lengths to $\lambda_{De}$. The physical parameters given above are also used as the initial state of nonlinear simulations presented in this paper, unless otherwise noted. The Python code is also used for solving the dispersion relation in a simple case of $T_e\rightarrow 0$, as shown in Fig.~\ref{fig:sim-gamma} where the limit of the full dispersion relation is presented for $k_z\lambda_{De}=0.005$.

For small values of the $k_z$ one can see in  Fig.~\ref{fig:gamma-kz} the fluid type reactive instability peaks near the resonant values of $k_y=m\cdot k_0$. For larger values of $k_z$, the fluid resonance is broadened by the thermal effects and for $k_z v_{Te} \geq \omega$, the resonance becomes kinetic, resulting in the smooth curve corresponding to the kinetic resonance instability of the ion sound in unmagnetized plasma \cite{GaryJPP1970a}.

The  regimes for different values of  $k_{z}$ can be seen from the dispersion equation (\ref{disp}). In the limit $k_{z}\rightarrow 0$, $\xi _{m}\gg 1$, and using $\mathrm{Z}\left( \xi _{m}\right)\rightarrow -1/\xi _{m}$,  the terms involving plasma dispersion functions result in cyclotron resonances:
\begin{gather}
\frac{\omega -\mathbf{k}\cdot \mathbf{v}_{0}}{\sqrt{2}k_{z}v_{e}}\mathrm{Z}%
\left( \xi _{m}\right) \rightarrow -\frac{\omega -\mathbf{k}\cdot \mathbf{v}_{0}}{\omega -\mathbf{k}\cdot \mathbf{v}_{0}+m\Omega _{ce}},\\
\xi _{m}=\frac{\omega -\mathbf{k}\cdot \mathbf{v}_{0}+m\Omega _{ce}}{\sqrt{2} k_{z}v_{e}}.
\end{gather}
This is the regime of the fluid (reactive) instability due to coupling of the ion sound and ion modes \cite{LashmorePhysA1970} which occurs for  cold plasma (and in the limit of  $k_z\rightarrow 0$). 

\begin{figure}[htp]
\includegraphics[width=0.485\columnwidth,clip]{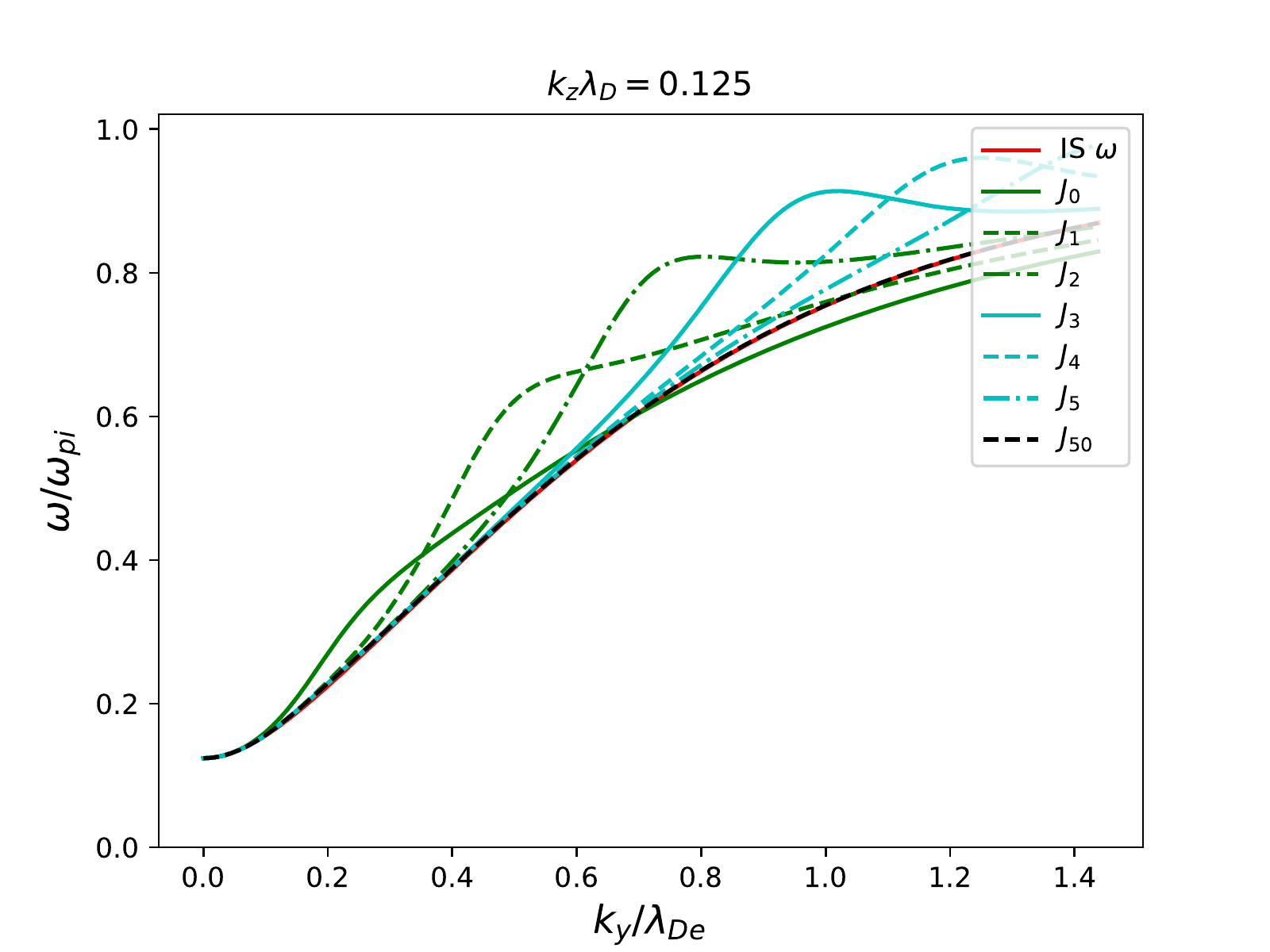}
\includegraphics[width=0.485\columnwidth,clip]{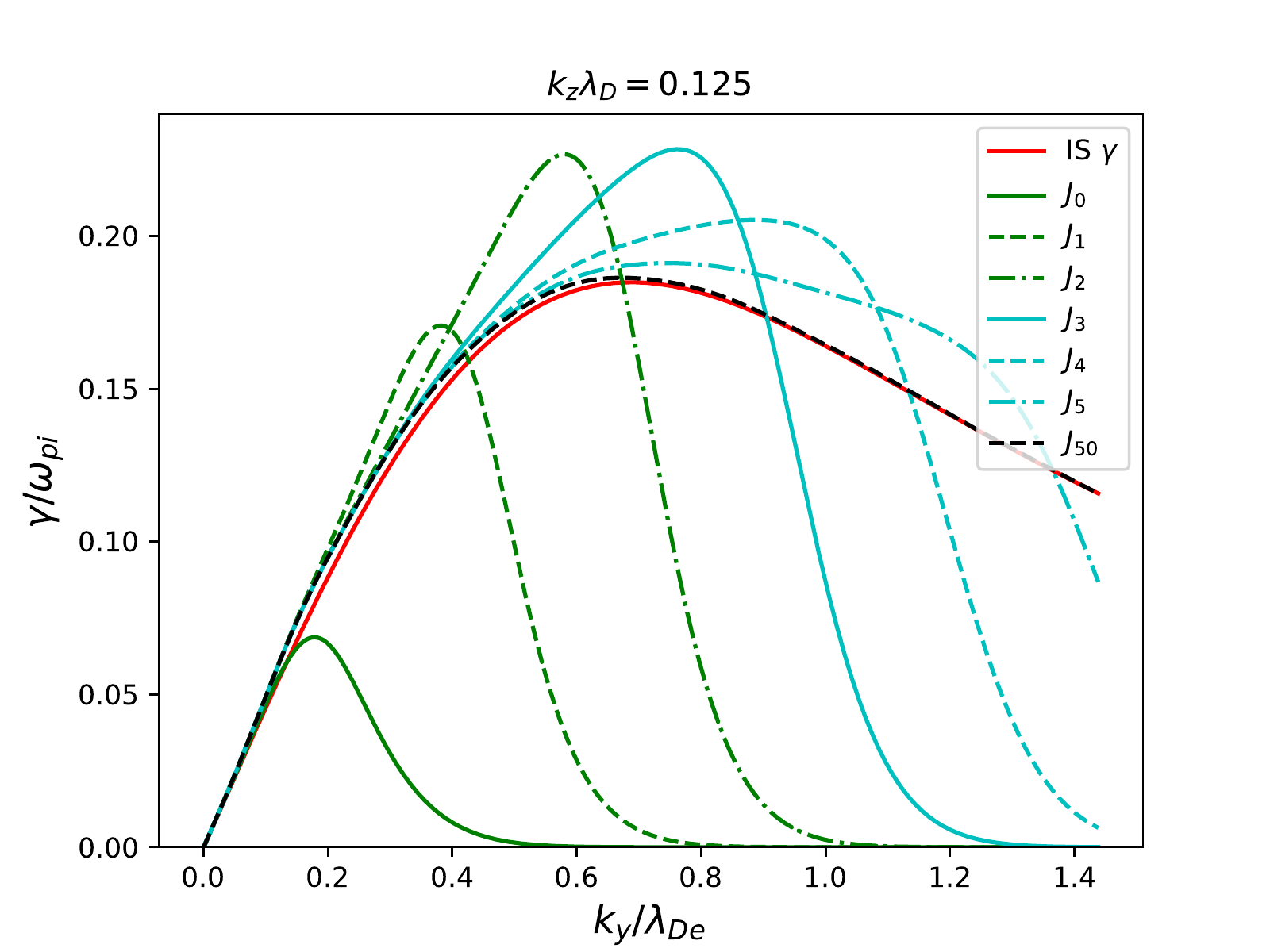}
\includegraphics[width=0.485\columnwidth,clip]{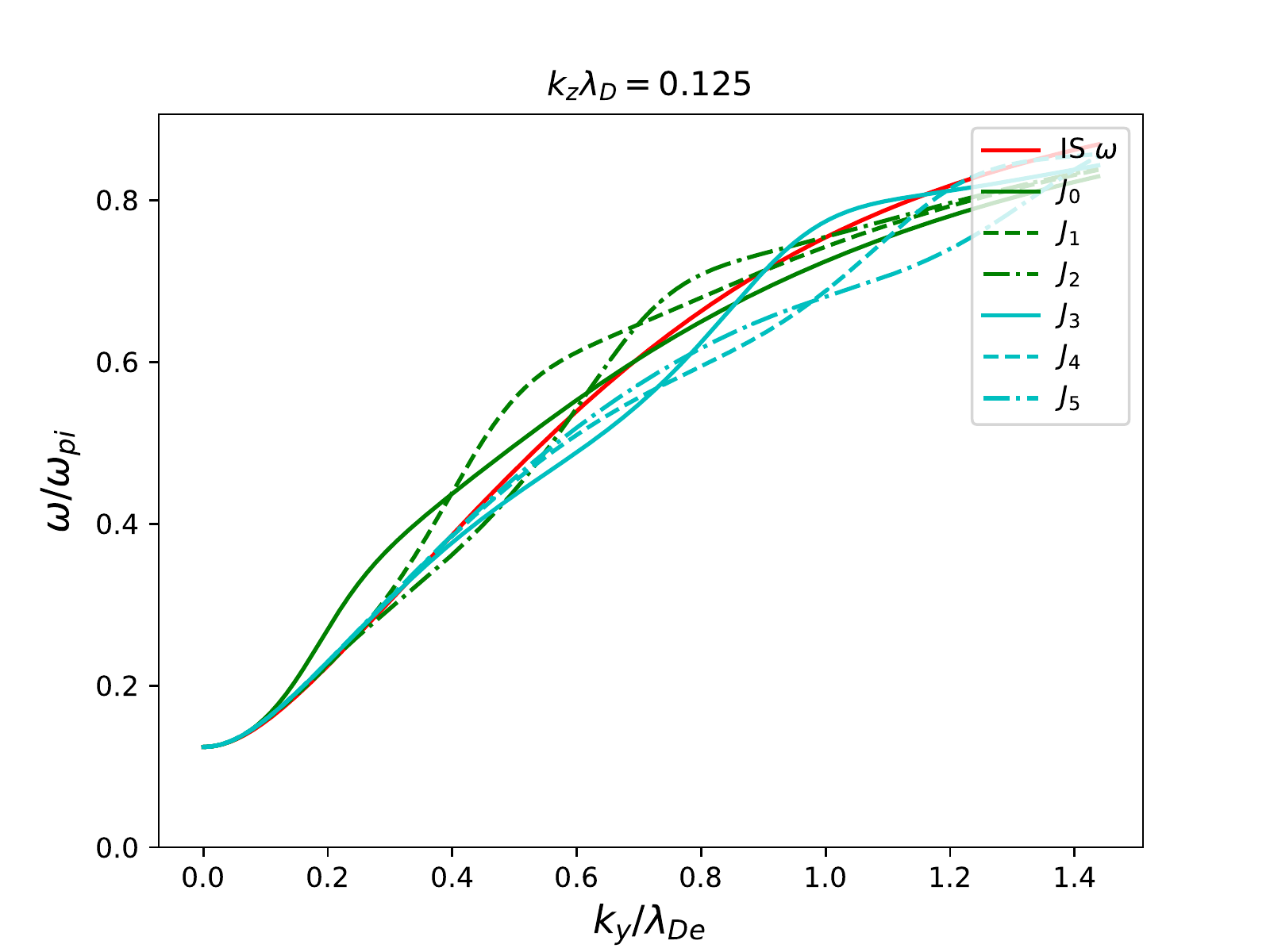}
\includegraphics[width=0.485\columnwidth,clip]{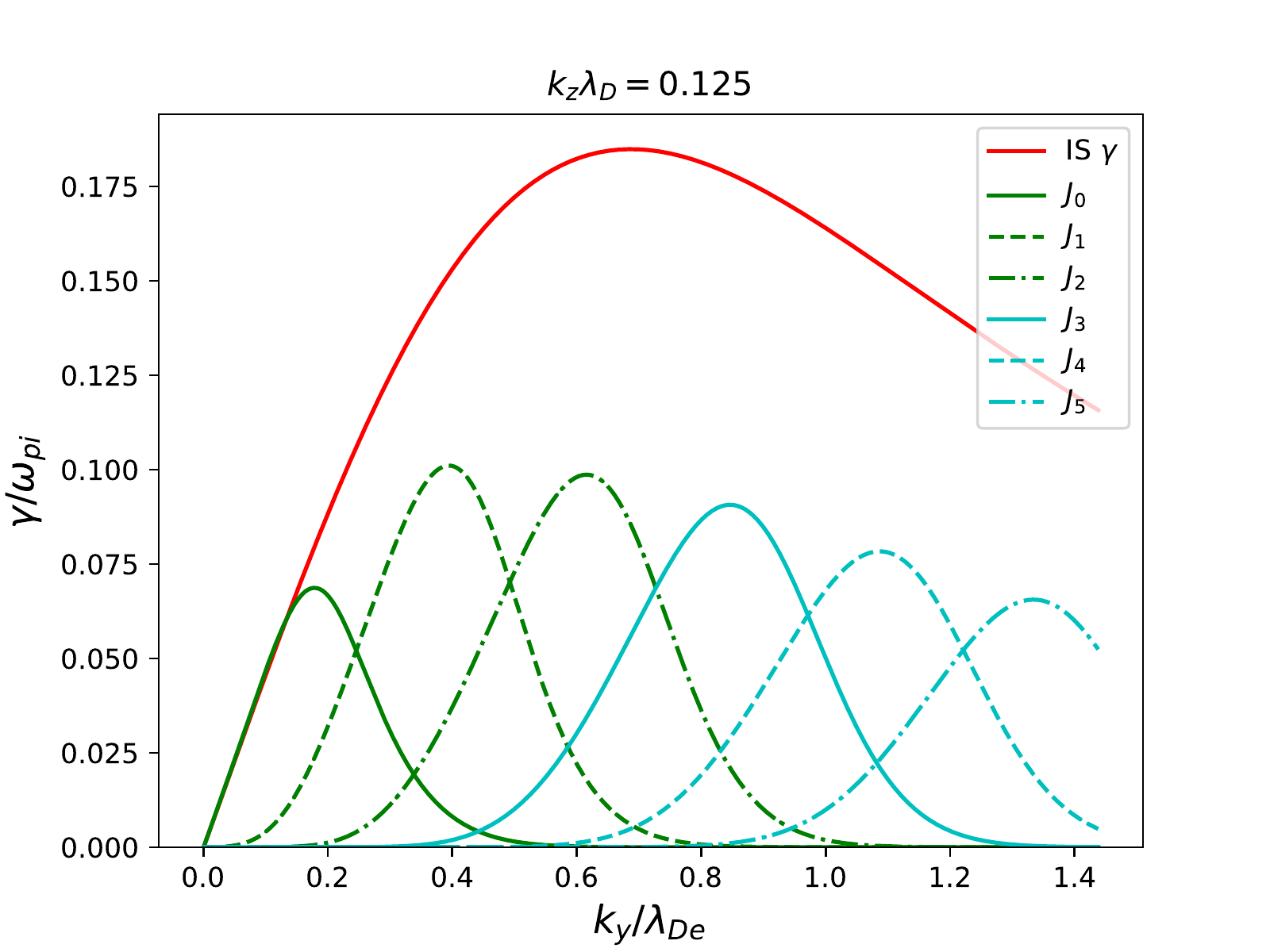}
\caption{Growth rate obtained from partial sums of the full dispersion relation (\ref{disp}); top row: terms up to $I_{\pm{k}}$ included solving for $\omega$ and $\gamma$; bottom row: only terms $I_{\pm{k}}$ of the sum included while solving for $\omega$ and $\gamma$. The solution from unmagnetized dispersion relation is included (labeled as IS); partial sum including terms up to $m=50$ is virtually indistinguishable from the unmagnetized solution. Parameters for this figure are: $v_0=-369\,c_s$, $v_{Te}=489\,c_s$, $\Omega_{ce}=96.5\,\omega_{pi}$.}\label{fig:gamma-resonances}
\end{figure}

 The thermal broadening of the  resonance and the transition to the kinetic ion sound instability  can be clearly seen in the example when only  $m=0$ is retained in the sum (\ref{el}):
\begin{equation}
1-\frac{\omega _{pi}^{2}}{\omega ^{2}}+\frac{1}{k^{2}\lambda _{De}^{2}}\left[
1+\frac{\omega -\mathbf{k}\cdot \mathbf{v}_{0}}{\sqrt{2}k_{z}v_{e}}e^{-b}%
\mathrm{I}_{0}\left( b\right) \mathrm{Z}\left( \frac{\omega -\mathbf{k}\cdot
\mathbf{v}_{0}}{\sqrt{2}k_{z}v_{e}}\right) \right] =0.\label{rede}
\end{equation}
The first three terms in this expression describe the ion sound mode (\ref{is}). Considering the last term as a small perturbation in Eq.~(\ref{rede}) one gets the so called modified ion sound instability \cite{ArefevTechPhys1970,ArefevNF1969}. Similar resonance broadening occurs for higher $m$ resonances, so that the resonant fluid type instability ($k_{z}\rightarrow 0$) is replaced by the kinetically driven mode for a finite $k_{z}$. This behavior is illustrated in Figs.~\ref{fig:gamma-kz} and \ref{fig:gamma-resonances}, where the mode frequencies and growth rates are calculated retaining only individual terms with different $m$ of the Bessel function series (lower panel) and partial sum up to the $m^\text{th}$ order (upper panel).  Note that the real part of the mode frequency for individual $m$ and partial sum $m$ modes always remain close to the ion sound mode frequency from Eq.~\ref{is}, see Fig.~\ref{fig:gamma-resonances}. For sufficiently large $k_{z}$, the summation of several components in the Bessel series illustrates the transition to unmagnetized ion-sound instability as shown in Fig.~\ref{fig:gamma-resonances}, where the solutions of the partial sum slowly converge to the unmagnetized curve as terms are added.

\section{Modified Two-stream Instability and Modified Two-Stream Buneman Instability}

The discussion in the previous section remains silent on one important feature of the instability with finite $k_z$: namely on the regime of the Modified-Two-Stream Instability (MTSI)\cite{McBridePF1972,LashmoreNF1973,Stepanov1965} where the effect of the electron parallel motion is typically considered in the context of the following dispersion equation\cite{LashmoreNF1973}
\begin{equation}
1-\frac{\omega _{pi}^{2}}{\omega ^{2}}-\frac{\omega _{pe}^{2}k_{z}^{2}}{(\omega -k_{y}v_{0})^{2}k^{2}}+\frac{\omega _{pe}^{2}k_{y}^{2}}{\Omega
_{ce}^{2}k^{2}}=0.  \label{mtsi}
\end{equation}

It is instructive to analyze the nature of this equation starting from the electron response in the form of Eq.~(\ref{ee}) which together with Eq.~\ref{ionc} results in the ion sound mode. The electron susceptibility in the form of Eq.~(\ref{ee}) is equivalent to the Boltzmann response for the perturbed electron density %
\begin{equation}
\widetilde{n}_{e}=\frac{e\widetilde{\phi }}{T_{e}}n_{0}  \label{b}.
\end{equation}
This expression follows from the parallel electron balance in neglect of the electron inertia
\begin{equation}
0=en\nabla _{\Vert }\widetilde{\phi }-T_{e}\nabla _{\Vert }\widetilde{n}.
\end{equation}

The electron inertia however can be neglected in the parallel momentum balance only when the condition  $\omega <k_{z}v_{e}$ is satisfied.  In presence of the strong transverse electron flow $v_{0}$,  apparent mode frequency  has to be modified due to the Doppler shift: $\omega \rightarrow \omega -k_{y}v_{0}$. When the condition $\omega -k_{y}v_{0} <k_{z}v_{e}$ is violated, the electron inertia terms have to be included. In this case, the electron continuity and momentum balance equations
\begin{gather}
-i\left( \omega -k_{y}v_{0}\right) \tilde{n}+ik_{z}n_{0}v_{\Vert}=0,\\
-im_{e}\left( \omega -k_{y}v_{0}\right) v_{\Vert}=iek_{z}\tilde{\phi},
\label{mb}
\end{gather}
result in the electron density response in the form
\begin{equation}
n_{e}=-\frac{ek_{z}^{2}\widetilde{\phi }}{m_{e}(\omega -k_{y}v_{0})^{2}}.
\label{nmsi}
\end{equation}
Note that when $\left( \omega -k_{y}v_{0}\right) \simeq k_{z}v_{e}$, the
assumption of the isothermal electrons becomes invalid, and the only
consistent approximation in the fluid theory is to assume $T_{e}=0$, so the
pressure term in (\ref{mb}) was omitted. 
In this equation, the third term corresponds to the electron density
perturbation from Eq.~(\ref{nmsi}) and the last term is due to the electron inertial perpendicular current. It is worth noting that the equation (\ref{mtsi}) is not fully consistent. 
The second term in this equation is obtained under ordering  $\left( \omega -k_{y}v_{0}\right) \simeq k_{z}v_{e}$, while the last term is obtained 
with the low
frequency approximation $\left(\omega -k_{y}v_{0}\right) \ll \Omega_{ce}$.
The latter may not be satisfied for some applications such as Hall thrusters. A more accurate dispersion equation is obtained from Eqs.~(\ref{disp}) and (\ref{el}) by taking a rigorous limit $T_{e}\rightarrow 0$ which yields the relation
\begin{equation}
1-\frac{\omega _{pi}^{2}}{\omega ^{2}}-\frac{\omega _{pe}^{2}k_{z}^{2}}{%
(\omega -k_{y}v_{0})^{2}k^{2}}-\frac{\omega _{pe}^{2}k_{y}^{2}}{((\omega
-k_{y}v_{0})^{2}-\Omega _{ce}^{2})k^{2}}=0,  \label{mbtsi}
\end{equation}%
which includes both the modified two-stream instability and the upper hybrid Buneman instability. In the following we will call this case the Modified Buneman Two-Stream Instability (MBTSI).

\begin{figure}[htp]
\includegraphics[width=\columnwidth,clip]{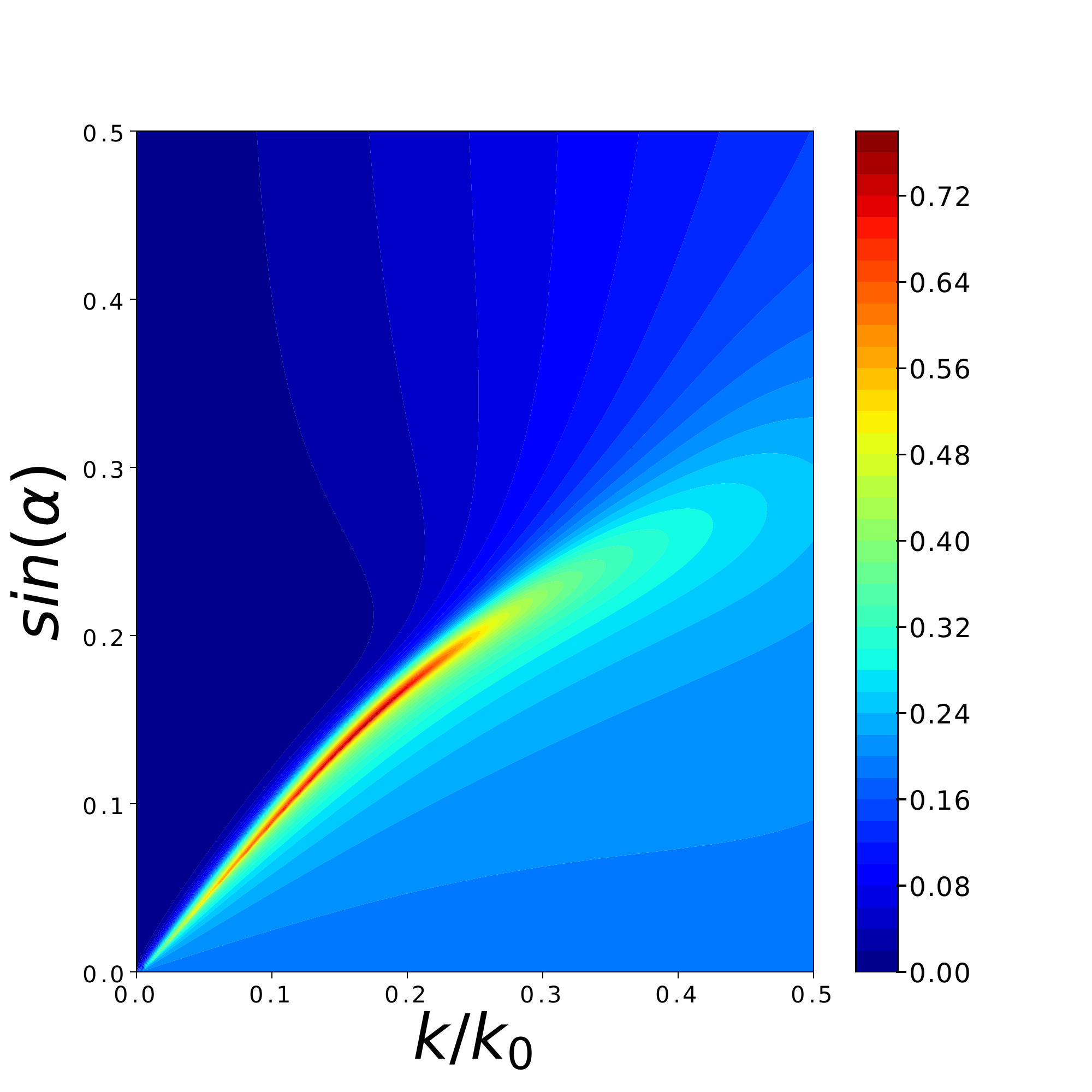}
\caption{Close-up of the modified two-stream instability (MTSI) growth rate $\gamma/\omega_{pi}$ given as a function of $\sin(\alpha)=k_z/k$ and $k / k_0$. Parameters for this figure are: $v_0=-369\,c_s$, $v_{Te}=489\,c_s$, $\Omega_{ce}=96.5\,\omega_{pi}$, $k_0\lambda_{De}=0.262$.}\label{fig:mtsi-full}
\end{figure}

The MBTSI regime is of particular importance as a finite value of $k_z$ result in the long wavelength instability at small $k_y/k_0 \ll 1$, well below the cyclotron resonances with $k_y\simeq m\cdot k_0$. This long wavelength mode, the leftmost peak in Fig.~\ref{fig:gamma-kz}, also shown in Fig.~\ref{fig:mtsi-full},  has a small growth rate, but as discussed in Section~\ref{sec:nonlinear}, turns out to be important in the nonlinear saturation regime enhancing the tendency toward long wavelength condensation. 

In our 2D simulations, which are bounded in the $z$-direction, it is observed generally that up to non-linear regime the instability grows uniformly everywhere, with $k_z \simeq 0$, and only after non-linear regime is reached does the wave like structure develop in the parallel direction. This can be explained by the fact that the linear growth rates increase monotonically towards $k_z\rightarrow 0$, while the the sheath allows fluctuations to extend up to the boundary by "insulating" the perturbations from the wall.   

\begin{figure}[htp]
\includegraphics[width=\columnwidth,clip]{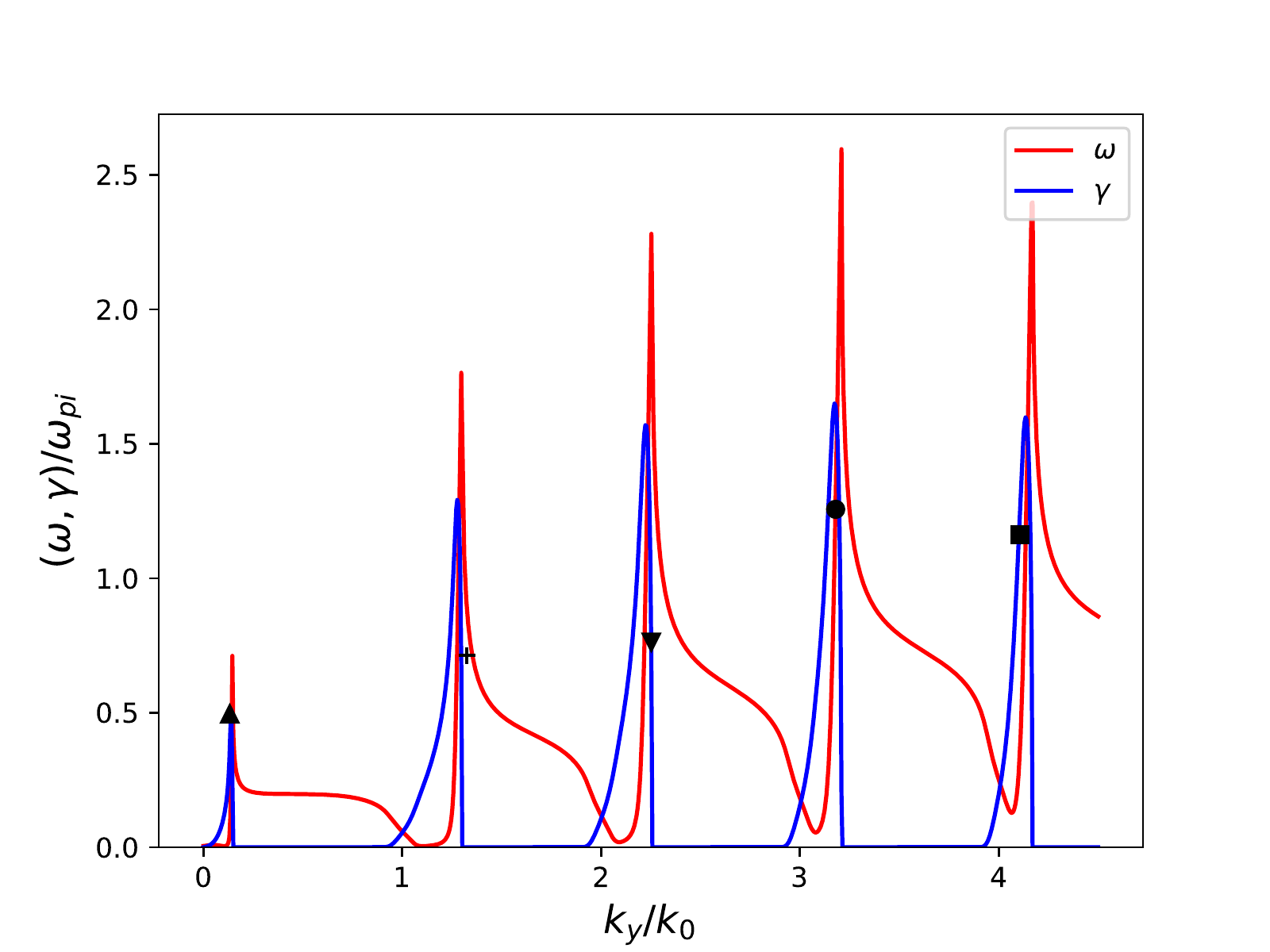}
\caption{Growth rate and frequency for ECDI instabilities for $k_z\lambda_{De}=0.00434$ at $T_e=10\,\text{eV}$. Growth rates as obtained from the 2D simulation (Fig.~\ref{fig:roots-sim}) are indicated by the symbols. Note the existence of the MTSI below the first cyclotron resonance. Refer to Table~\ref{table:rates} for the values. Parameters for this figure are: $v_0=-369\,c_s$, $v_{Te}=489\,c_s$, $\Omega_{ce}=96.5\,\omega_{pi}$, $k_0\lambda_{De}=0.262$.}\label{fig:omega-gamma}
\end{figure}

Generally in the literature the ECDI instability was considered\cite{CavalierPoP2013} for typical values  of the parallel wave length of $k_z \lambda_{De}\simeq 0.01-0.09$ corresponding to a parallel wave length of $\lambda=7.4-0.8\,$cm, for typical Hall-effect thruster parameters. It could be misleading, however, to estimate $k_z$ and relevant dynamic regimes of the ECDI based on full wave lengths in a Hall-effect thruster gap. It has been noted earlier\cite{ChenJNE1965,ChenPF1965,ChenPF1979,BarrettPRL1972} that in a bounded plasma the sheath effects allow the wavelength along the magnetic field to be much longer than would otherwise follow from the the geometrical constraint $k_z=2\pi/L_r$, where $L_r$ is the plasma width between the boundaries. In our 2D simulations we observe that the parallel structure along the magnetic field is consistent with a half-wavelength fit between the boundaries. This corresponds to $k_z\lambda_{De}=0.00434$. We used this value for calculations of the linear growth rates in Figs.~\ref{fig:sim-gamma} and \ref{fig:omega-gamma}.

\section{Growth and saturation of the cyclotron harmonics of ECDI and MTSI modes}

PIC2D is a 2D3V PIC code developed by Dmytro Sydorenko at University of 
Alberta based on his earlier 1D3V PIC code, EDIPIC.\cite{SydorenkoTh2006}
The initial state of the PIC2D simulation is a $v_0$-shifted Maxwellian distribution for the electrons, and a stationary Maxwellian for the ions. The initial state is quasi-neutral and homogeneous, equipped with the typical Hall-effect thruster parameters presented in the previous section (and Appendix). The magnetic field lines are terminated at a dielectric boundary, so a sheath potential develops soon after time starts running. The simulation box is $L_r=53.8\,\text{mm}$ by $l_\theta=13.45\,\text{mm}$, and the numerical parameters are detailed in Table~\ref{table:numerical_params} in Appendix~\ref{app:determine}. Even though to code allows to do so, no collisions were used in our simulations.

\begin{figure}[htp]
  \begin{tikzpicture}
    \node (img1) {\includegraphics[width=0.95\columnwidth,clip]{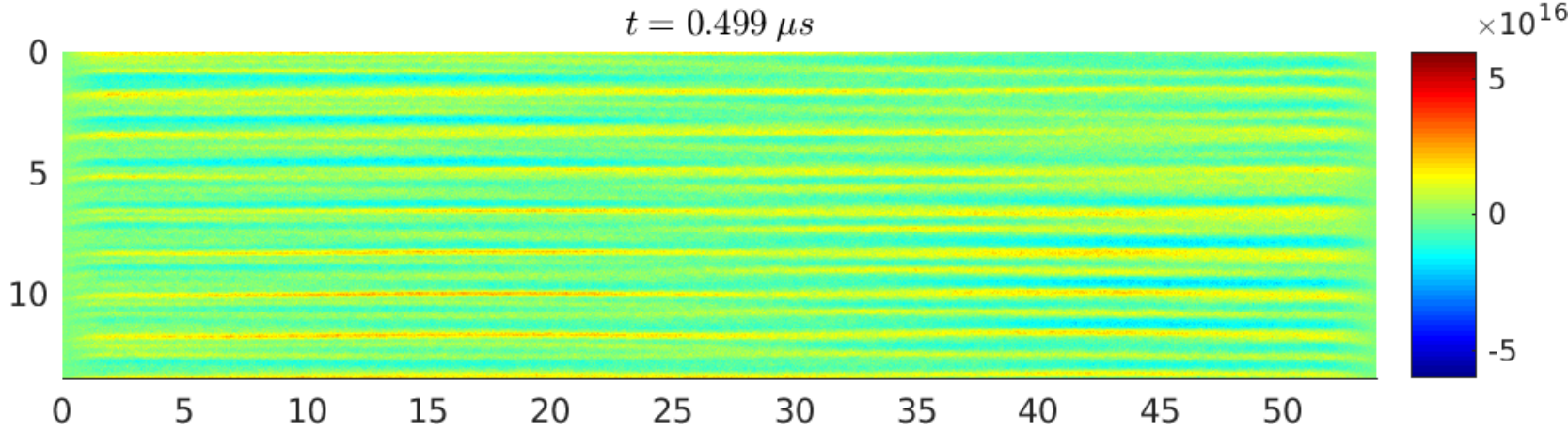}};
    \node[below of=img1,node distance=1.75cm] (img2) {\includegraphics[width=0.95\columnwidth,clip]{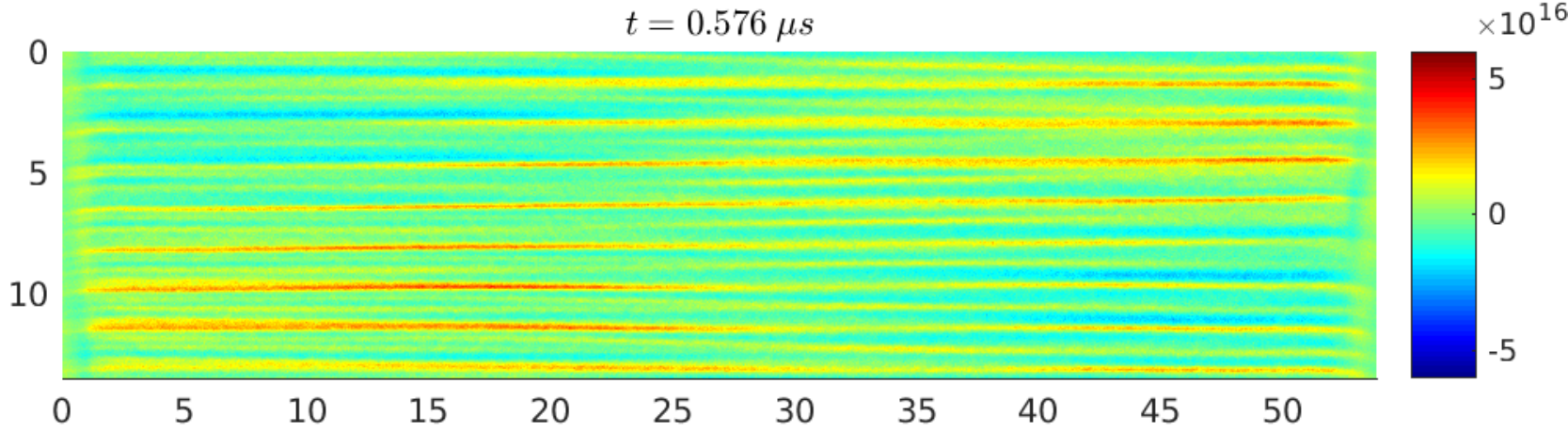}};
    \node[below of=img2,node distance=1.75cm] (img3) {\includegraphics[width=0.95\columnwidth,clip]{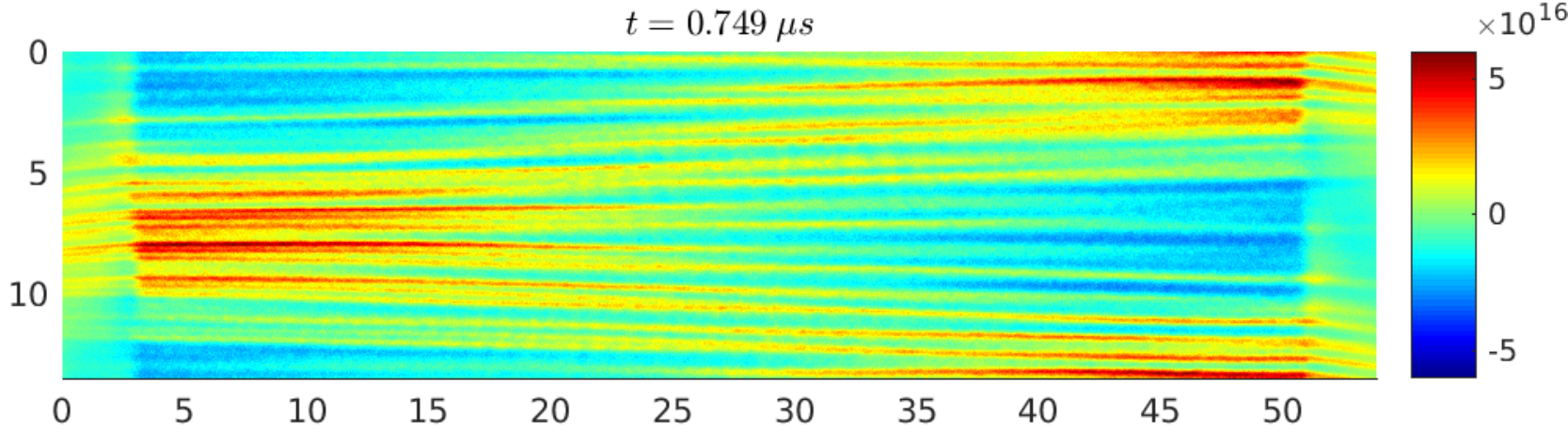}};
    \node[below of=img3,node distance=1.75cm] (img4) {\includegraphics[width=0.95\columnwidth,clip]{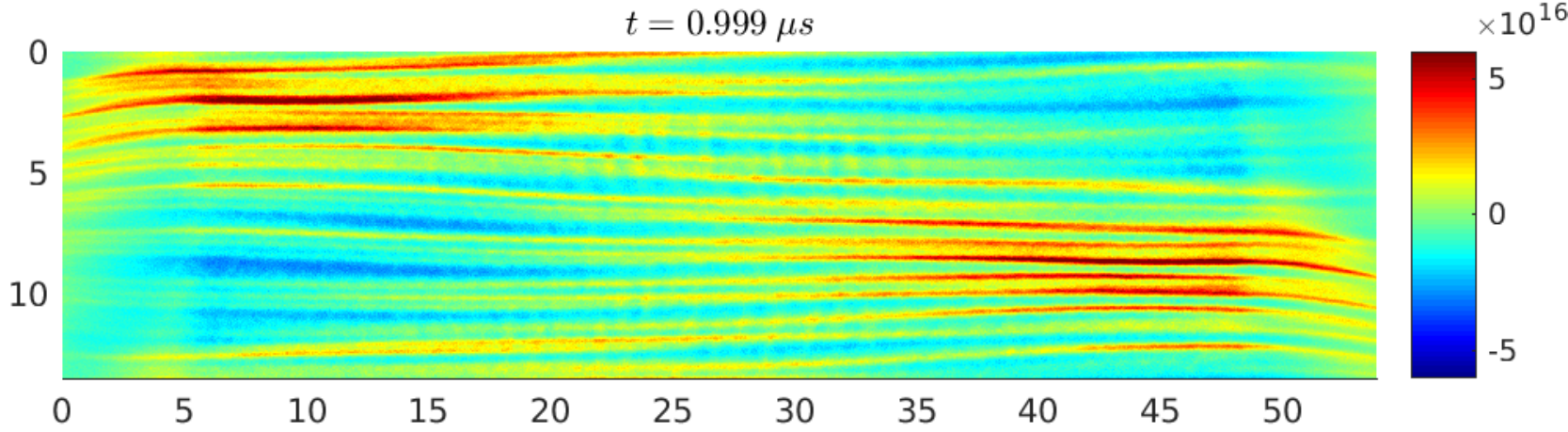}};
    \node[below of=img4,node distance=1.75cm] (img5) {\includegraphics[width=0.95\columnwidth,clip]{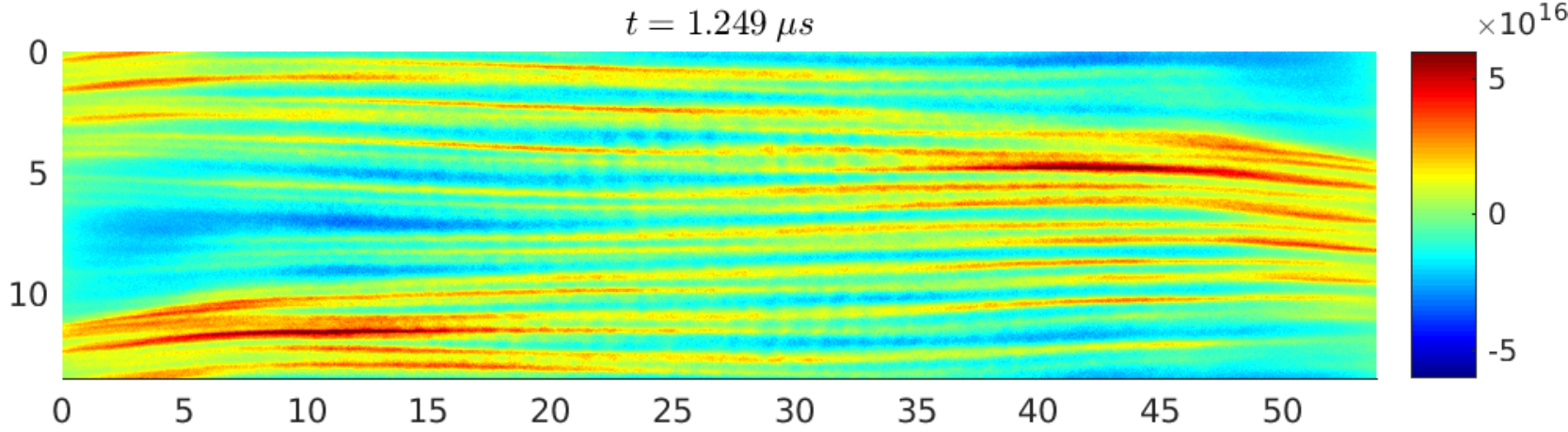}};
    \node[below of=img5,node distance=1.75cm] (img6) {\includegraphics[width=0.95\columnwidth,clip]{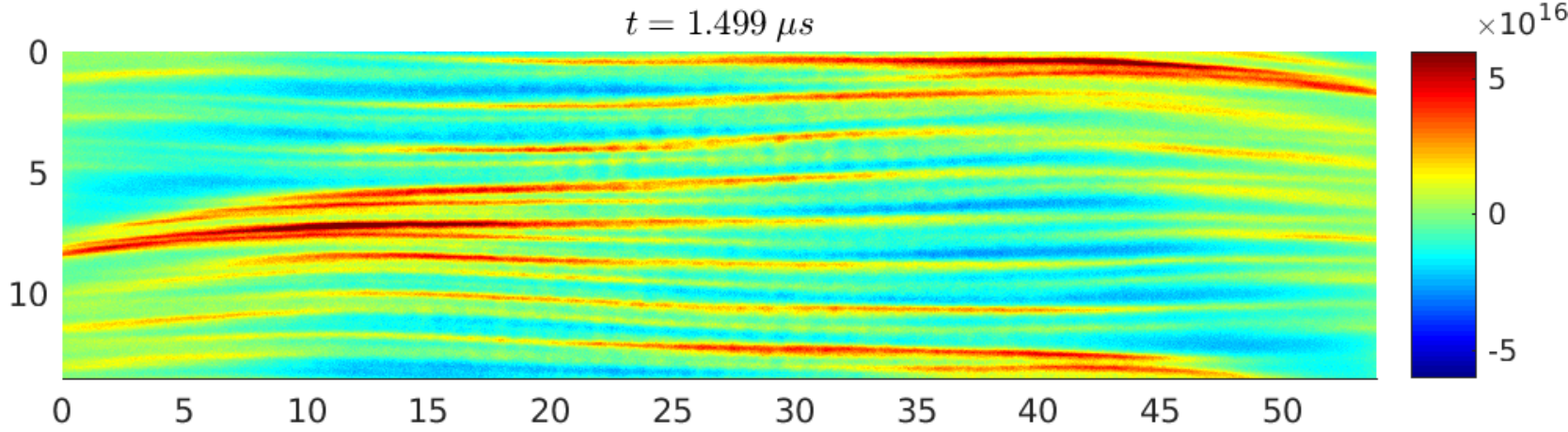}};
    \node[below of=img6,node distance=1.75cm] (img7) {\includegraphics[width=0.95\columnwidth,clip]{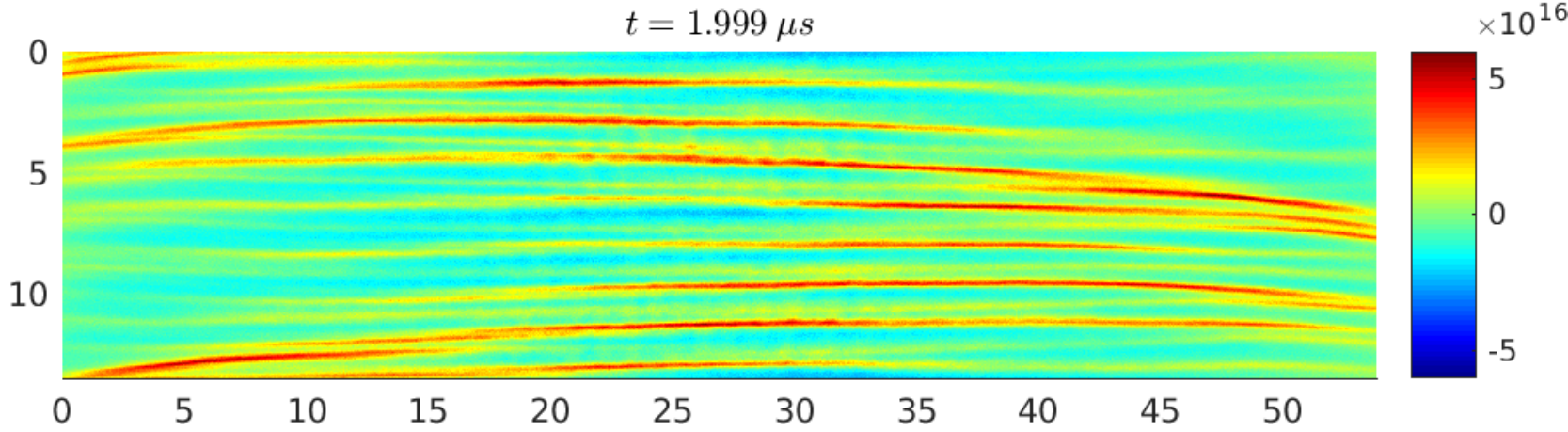}};
    \node[below of=img7,node distance=1.4cm]{$z~/~\text{mm}$};
    \node[left of=img4,rotate=90,yshift=3.5cm]{$y~/~\text{mm}$};
  \end{tikzpicture}
\caption{Time slices of the ion density fluctuation $n_i-\langle n_i\rangle_y(z,t)$, showing growth and saturation of the predominant $m=1$ cyclotron mode followed by the development of structure parallel to the magnetic field.}\label{fig:stages}
\end{figure}

\begin{figure}[htp]
\includegraphics[width=\columnwidth,clip]{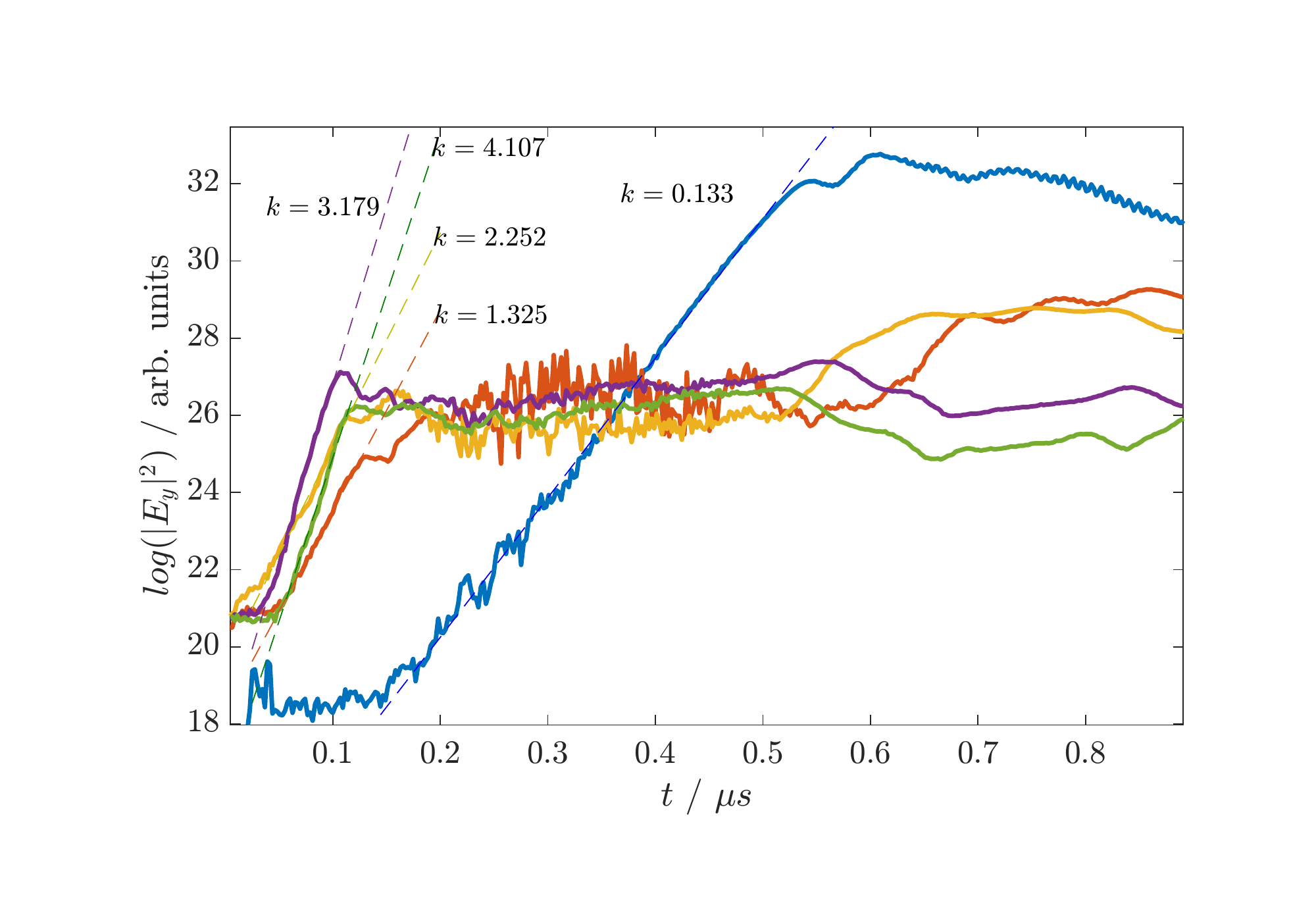}
\caption{Logarithmic amplitude of the azimuthal electric field energy $|E_y|^2$, showing stages of linear growth and early nonlinear saturation of the cyclotron and MTSI modes. Linear least-squares fitting has been used to obtain values of the growth rates for each mode (labeled by their $k$ values wrt/$k_0$), given in Table~\ref{table:rates}.}\label{fig:roots-sim}
\end{figure}

Chronologically, the evolution of the simulations goes as follows (see figures~\ref{fig:stages} and \ref{fig:roots-sim}). First, in the linear-like stage the fastest growing modes -- the ECDI with $m=3-5$ -- grow and introduce some heating to the electrons primarily in the perpendicular direction.  Due to heating and nonlinearities the lowest $m=1$ resonance becomes the ECDI mode with the largest energy content, even though linearly it is the slowest growing mode. The resonance condition $k_y=m \Omega_{ce}/v_0$ gives, with our choice of $l_\theta=13.45\,\text{mm}$ the quantum numbers $n_y=m\cdot 7.534$, but due to kinetic effects the maximum growth rates are found at higher values of $n_y$. It is observed consistently with the linear dispersion relation that the maximum growth rates for the $m$ cyclotron resonances correspond to $n_y (m=1,2,3..)=\{10,17,24,31,...\}$ as ECDI modes. For higher resonances the up-shift in $k_y$ is lower, diminishing the gap between maximum growth rates of $m$-resonances to ${\Delta}n_y=7$. The mode amplitudes are shown in Fig.~\ref{fig:stages}, and growth rates are given in Table~\ref{table:rates}. As can be also seen from Fig.~\ref{fig:omega-gamma}, these locations are very close to the maximum growth rates obtained from the linear dispersion relation. After the $m=1$ and $m=2$ modes saturate, the MTSI mode starts growing, suggesting non-linear feedback between the modes. During the growth of the MTSI, mode competition between $m=1$ and $m=2$ ECDI modes is observed. The MTSI mode grows, heating the electrons predominantly in the parallel direction due to the parallel electric field. Heating in parallel direction  results in enhanced losses to the sheath and saturation of the MTSI. The ECDI modes stay at their saturated level that was established earlier, but after the saturation of the MTSI the $m=1\,\&\,2$ modes grow by 5-10 \%, while $m=3\,\&\,4$ modes lose energy correspondingly. At this stage, we observe the saturation of the anomalous axial current, as shown in Fig.~\ref{fig:current-total}.

Linear growth rates may be determined directly from simulation data when the runs are performed with good enough resolution. The procedure for finding the linear growth rates from the spectrogram of a simulation is outlined in Appedix~\ref{app:determine}. The linear and early non-linear stage of the evolution of individual mode resonances is shown in figure~\ref{fig:roots-sim}, where we also provide the values for normalized wave number from the linear fits that can be made to the modes. The values obtained from simulations are given in table~\ref{table:rates}, and also plotted for the comparison with the linear dispersion relation solutions in  figure~\ref{fig:omega-gamma}. As can be seen, the growth rates are of the right magnitude, although depressed due to the short time of growth available for fitting. The MTSI mode (first peak) is well represented though, showing the importance of good statistics. 
The MTSI peak is well represented because it grows to a higher amplitude, giving a larger range for least-squares fitting. This explains why the ECDI growth rates are generally slightly underestimated by the 2D simulations; the mode energy for the ECDI modes has only four periods of growth (before nonlinear stage), whereas the MTSI has more than ten periods for fitting (see Fig.~\ref{fig:roots-sim}). Part of the ECDI growth curve is affected by early nonlinear saturation processes, decreasing the apparent growth rate.

To emphasize this point, we ran a case for $T_{e}\approx 0$ with the 2D code, and a case with 1D version of the code using very good statistics (Fig.~\ref{fig:sim-gamma}) that gives a remarkably good value for both the growth rate and frequency of the Modified Buneman Two-Stream instability (equation~\ref{mbtsi}).

\begin{table}[htp]
\begin{tabular}{|c|c|c|c|}\hline
Symbol & $n_y$& $k/k_0$&$\gamma / \omega_{pi}$\\\hline
$\blacktriangle$ &1 &0.1325&0.4957\\
\ding{58} &10 &1.3248&0.7133\\
$\blacktriangledown$ &17 &2.2521&0.7629\\
$\bullet$ &24 &3.1794&1.2568\\
$\blacksquare$ &31 &4.1068&1.1628\\\hline
\end{tabular}
\caption{The wave length and growth rate of unstable modes observed in  the linear stage simulations as shown in Figs.~\ref{fig:roots-sim} and \ref{fig:omega-gamma}.}\label{table:rates}
\end{table}

A curious feature of the ECDI/MTSI energy in Fig.~\ref{fig:roots-sim} is how the growth of the MTSI commences only after the ECDI modes have saturated, and how the energy of the ECDI modes remains relatively constant while the MTSI is growing. This emphasizes that the modes are not in fact independent, but are nonlinearly coupled instead. It is observed that in the sheath-bounded radial direction the mode assumes a half-wave pattern with $n_z=1/2$, whereas in the azimuthal (periodic) direction the mode is a full wave with $n_y=1$. If we use this observation to prescribe $k_z=\pi/L_r$, we get Fig.~\ref{fig:omega-gamma} from the dispersion relation, and our MTSI peak aligns well with the peak of the MTSI in $k_y=2\pi/l_{\theta}$, and the fit to ECDI modes with $n_y=\{10,17,24,31,...\}$ is also satisfactory.

There are 3D simulations that indicate the existence of long-wavelength modes very similar to those found in our simulations\cite{Taccogna}, suggesting that the presence of the MTSI is a robust feature even in full Hall-effect thruster geometry.

\section{Nonlinear spectra of ion and electron density: short wavelength features}\label{sec:nonlinear}

We have earlier noted that the cyclotron resonances drive strongly coherent cnoidal type waves which are limited by saturation through ion dynamics\cite{JanhunenPOP2018}, while the waves retain their cyclotron resonance characteristics far into the simulation even with significant electron heating and nonlinear interactions between modes.

In 2D the dispersion relation admits a long wavelength instability that is absent in 1D, namely the modified two-stream instability of Section~\ref{sec:instability}. The latter mode is observed to modify the nonlinear dynamics compared to the 1D case, resulting in significantly faster evolution of the long wavelength components, and a lower energy content (or, fluctuation level) is retained in the ECDI modes. 

\begin{figure}[htp]
\includegraphics[width=0.8\columnwidth,viewport=45 89 719 507,clip]{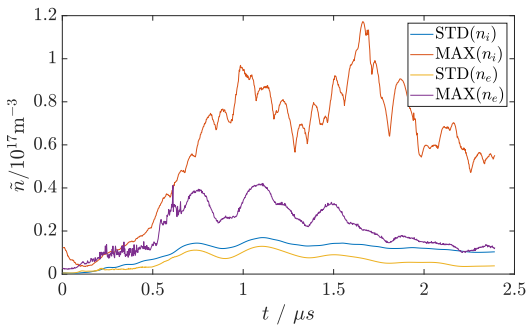}
\caption{Ion and electron density fluctuation levels over the whole simulation volume; standard deviation and maximum.}\label{fig:delta-n}
\end{figure}

\begin{figure}[htp]
\includegraphics[width=\columnwidth,clip]{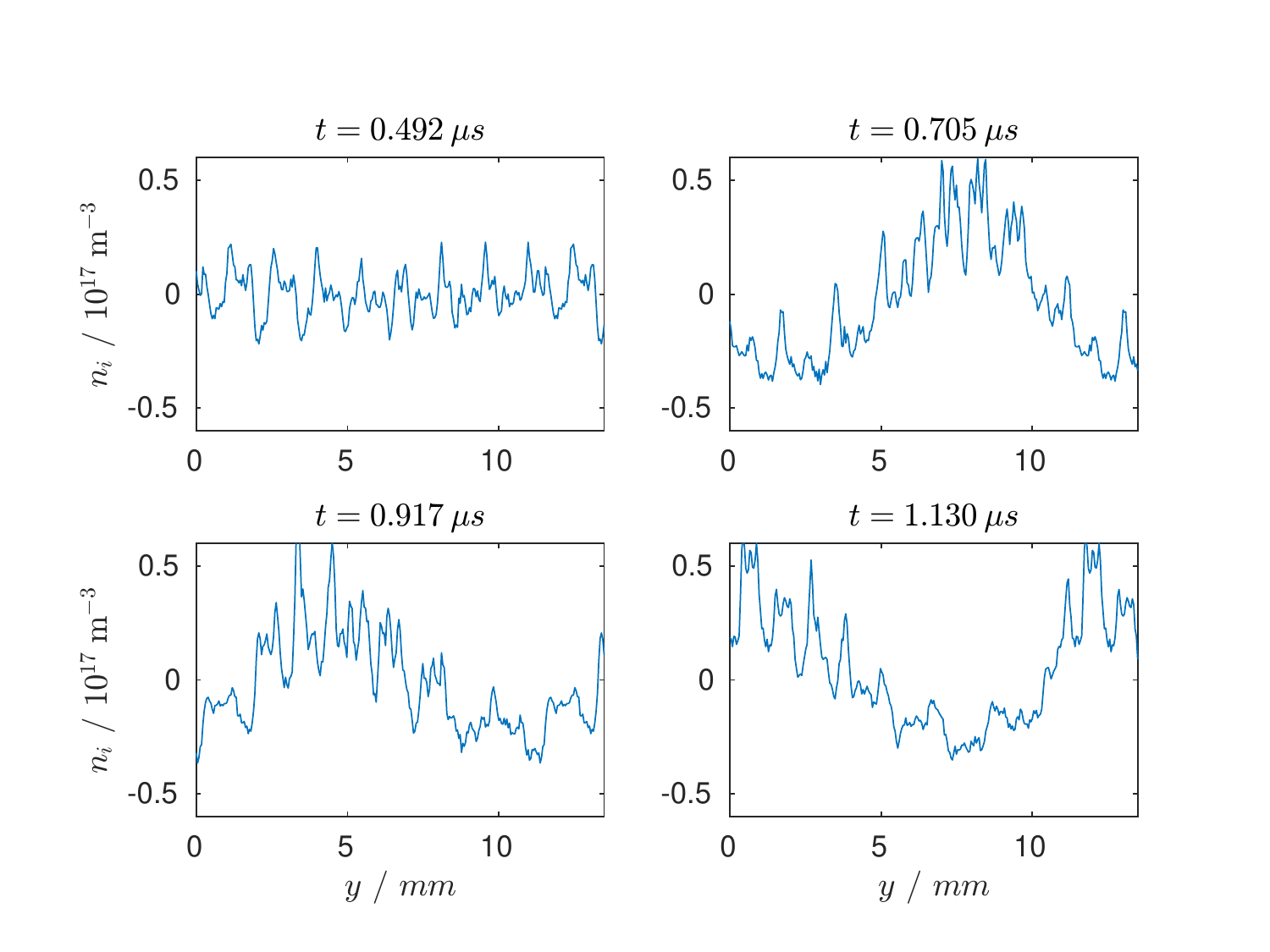}
z\caption{Stages of the nonlinear development of the ion density fluctuations at four time slices. Large scale mode is formed, the wave crest compressing the EC waves, merging of the peaks with shift to lower $k$ and reshaping of the wave packet to a more triangular form. Plots are from $r=5\,\text{mm}$.}\label{fig:condensation-slices}
\end{figure}

\begin{figure}[htp]
\includegraphics[width=0.95\columnwidth,clip]{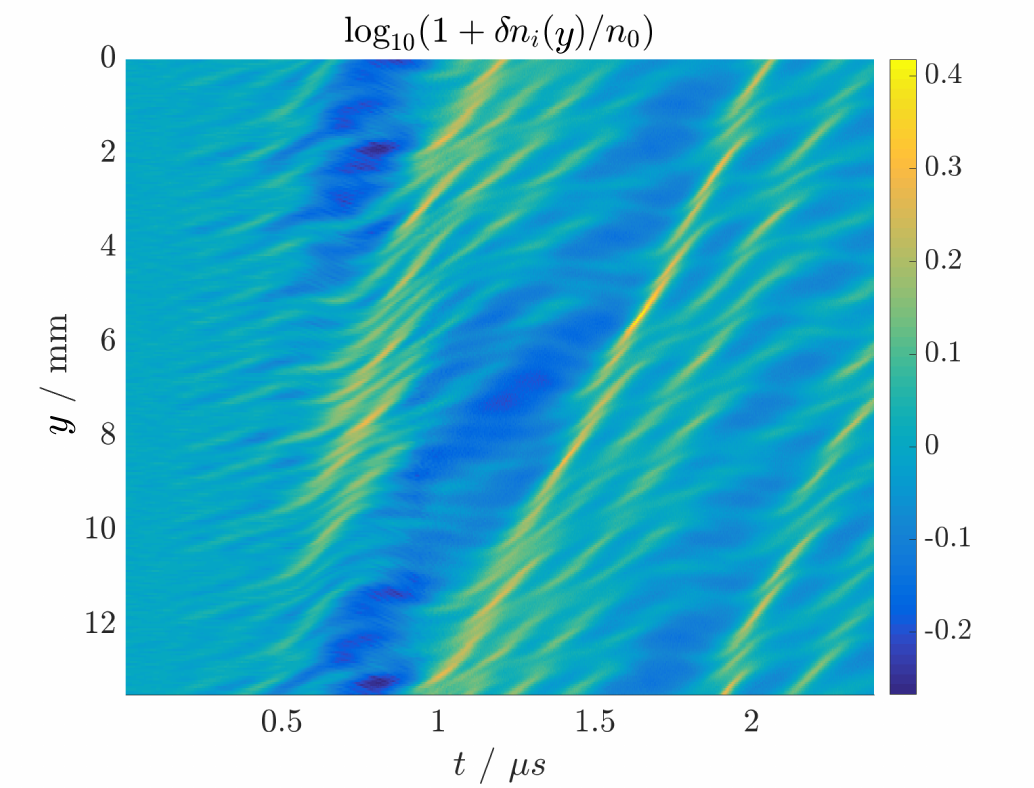}
\includegraphics[width=0.95\columnwidth,clip]{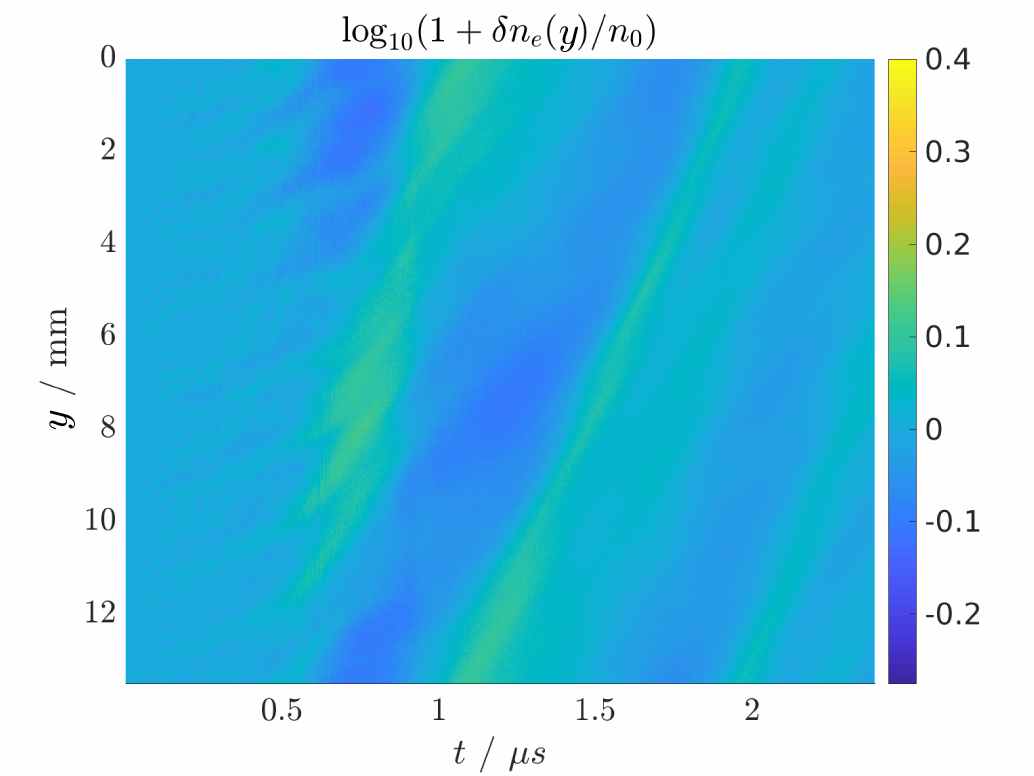}
\caption{Top: Azimuthal ion density fluctuations as a function of time. Individual wave crests propagate at different phase velocities with respect to the wave packet, exchanging energy with one another. Bottom: Azimuthal electron density fluctuations as a function of time. After saturation, electron density assumes a smooth profile. Volume-averaged root-mean-square fluctuation level is roughly half of the ion fluctuation level. The figures are scaled the same way, $n_0=10^{17}\,\text{m}^{-3}$ to fix units although $\delta\hspace{-0.25ex}n=n-\langle n\rangle_y(z,t)$, and both are measured at $z=5\,\text{mm}$.}\label{fig:azim-sec}
\end{figure}

After the early nonlinear stage, the convective nonlinearity compresses/expands the ECDI in the hills/troughs of the now-dominant MTSI mode (figure \ref{fig:condensation-slices}), causing jet-like injection to the sheath from the compressed maxima (particularly evident in the 4th figure of Fig.~\ref{fig:stages}) with accompanying faster decay of the plasma profile. Transient large fluctuations in the electron density predominantly in the parallel direction are observed to originate from the sheath at this stage, also seen as parallel ripples in the ion density, which likely act as a relaxation mechanism. At this stage (around $1\,\mu\text{s}$ into the simulation) the linear mode characteristics become a poor descriptor of the system: electron density fluctuations do not significantly increase, but ion density fluctuations grow (see figures~\ref{fig:delta-n}-\ref{fig:azim-sec}), and group velocity of the wave packet increases significantly. Feedback between the ECDI- and MTSI-scale modes is apparent, with expansion and compression of the wave crests as the wave packet progresses. 

An important feature of the nonlinear dynamics observed in our 2D simulation is the the difference in the behavior of the ion and electron density. After heating, electron fluctuations become fairly uninteresting (and low-amplitude), but the ion density exhibits a wealth of non-linear phenomena. This difference is especially apparent in the short wavelength $k_y\lambda _{De}<1$ part of the spectrum. The ECDI still remains the dominant mode of energy injection into the short wavelength ion-sound fluctuations, whose frequency approach $\omega_{pi}$ in this limit, but strongly modified by signatures of nonlinear wave breaking \cite{Davidson_Nonlinear_methods} due to ion dynamics. In the strong turbulence state, the ion density fluctuations in the azimuthal direction become cnoidal-like, and are lead by a wave with a very sharp peak of positive amplitude, after which a train of crests of decreasing amplitude follow. As apparent from figure~\ref{fig:azim-sec}, where a radial section of the plasma at $5\,\text{mm}$ is shown, the crests do not propagate as much as exchange energy through elastic-like collisions (akin to soliton collisions), so the amplitude maximum travels at a higher speed than the individual crests do, as may be the case for envelope solitons\cite{Remoissenet2013Waves}. Similar features were observed in 1D simulations as well, although realized after a significantly longer time of simulation.

\section{Parallel electron heating due to MTSI}

A new feature with respect to our earlier 1D simulations is the rapid parallel heating in the 2D simulations, observed to occur in the same pattern as the MTSI mode. It is well-known that the MTSI is an effective heating mechanism for electrons along the magnetic field \cite{McBridePF1972} and the heating is the likely saturation mechanism for the MTSI, as larger parallel temperature will induce large losses of high energy electrons into the sheath.
\begin{figure}[htp]
\includegraphics[width=\columnwidth,clip]{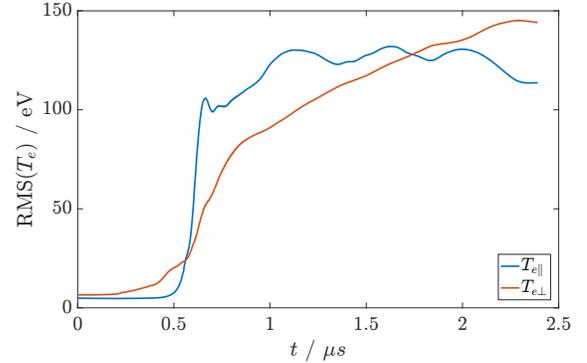}
\caption{Anisotropic  heating of the $T_{e\perp}$ and $T_{e\parallel}$ components. Parallel heating is due to the MTSI mode. The curves show the volume averaged parallel and perpendicular temperatures.}\label{fig:heating}
\end{figure}
\begin{figure}[htp]
\includegraphics[width=\columnwidth,clip]{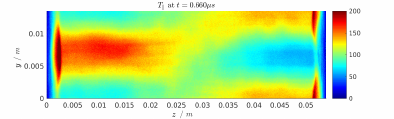}
\includegraphics[width=\columnwidth,clip]{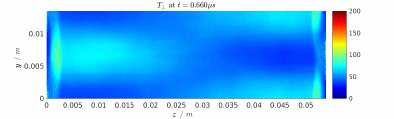}
\caption{Spatial structure of the perpendicular $T_{e\perp}$ and parallel $T_{e\parallel}$ electron temperatures.}\label{fig:heating-structure}
\end{figure}

After saturation of parallel temperature  $T_{e\parallel}$ , electron heating in the perpendicular direction catches up to parallel heating (which has saturated through sheath losses), after which the state of strong turbulence is reached (see Fig.~\ref{fig:heating}). Even in this stage, the ECDI mode structure is prominent in density spectra, and is clearly affected into a ``wave street'' (alternating maxima and minima) characteristic to the MTSI mode, although the large scale structure is less apparent. These features are unchanged in simulations with half time-step, ruling out the possibility of numerical heating. The heating profile is shown in Fig.~\ref{fig:heating-structure}, clearly indicating the MTSI mode as the cause of parallel heating (and to a lesser degree, perpendicular). The growth of the MTSI mode terminates when the $T_{e\parallel}$ growth terminates in Fig.~\ref{fig:roots-sim}.

\section{Spectral cascade down to long wave lengths}

\begin{figure}[htp]
\includegraphics[width=\columnwidth,viewport=63 105 700 486,clip]{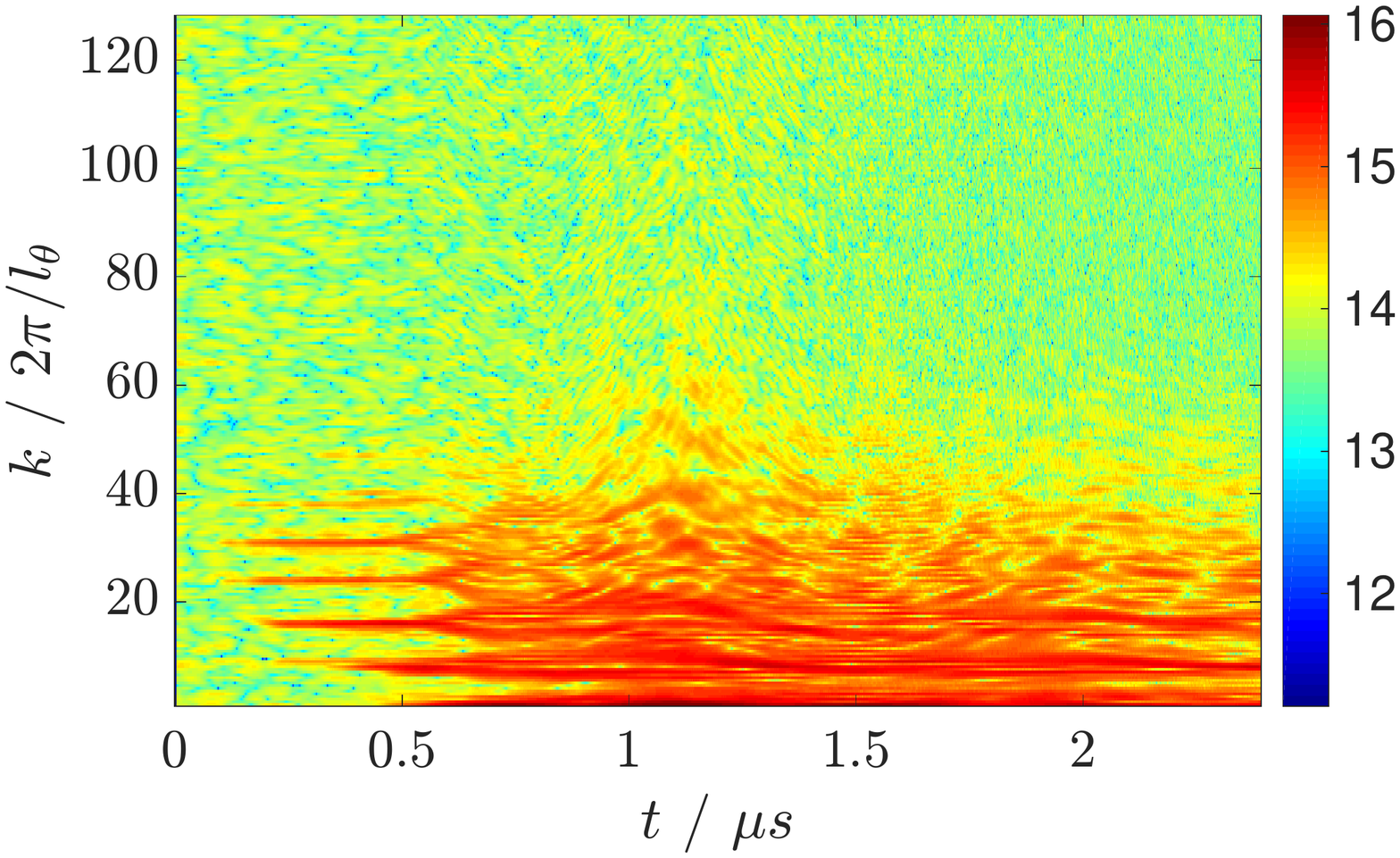}
\includegraphics[width=\columnwidth,viewport=45 80 720 496,clip]{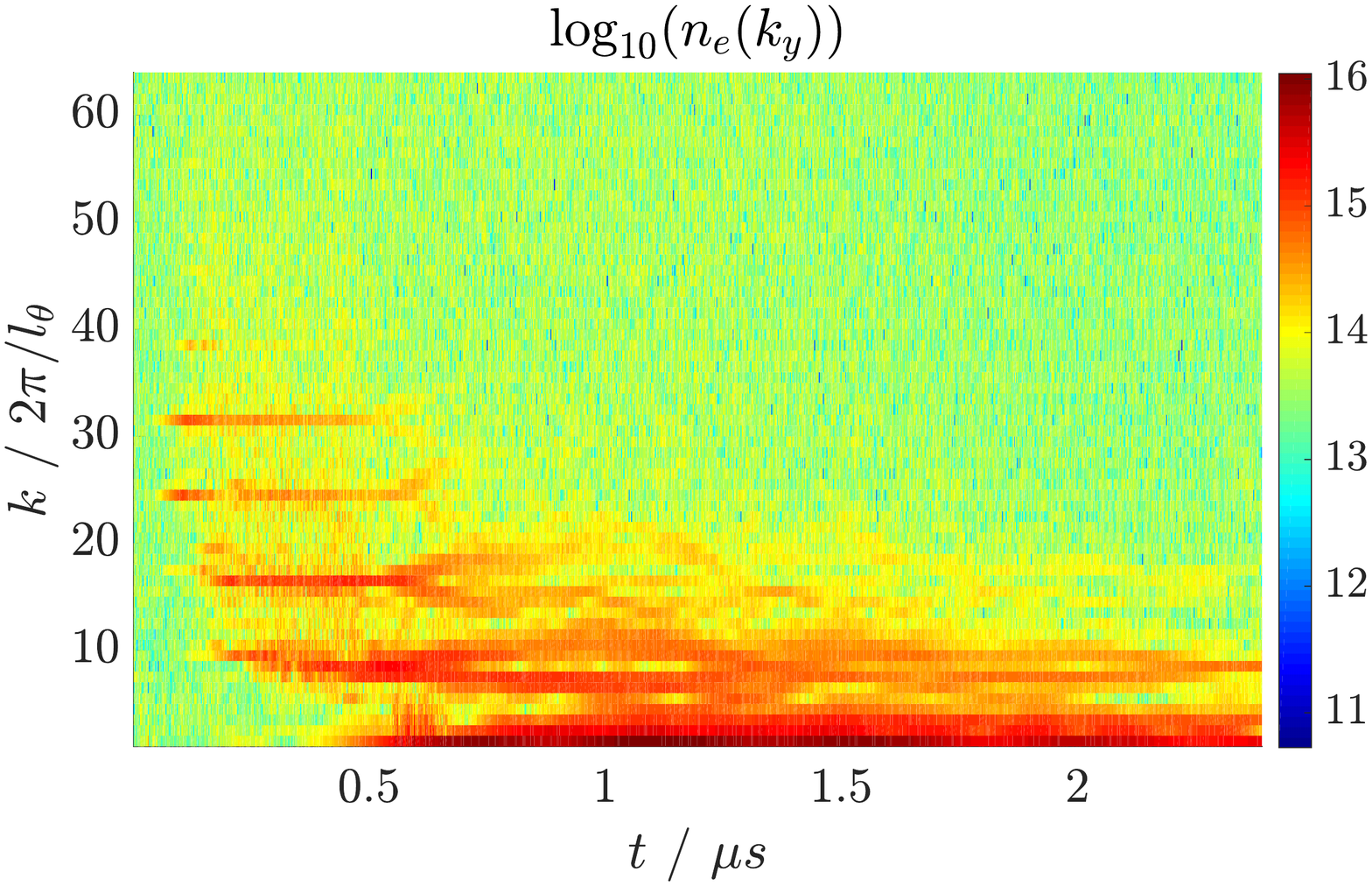}
\caption{Evolution of the azimuthal ion density (top) and electron density (bottom) $k$-spectra over time at $L/4$ of the simulation, plotted as $\log_{10}{\tilde{n}_{i,e}(k_y)}$. Discrete peaks occur at the $\omega-k v_0-=m\Omega_{ce}$ resonances, and the lowest peak (after $0.5\mu\text{s}$) is the MTSI mode around $k_y=2\pi/l_\theta$, or $n_y=1$.}\label{fig:ion-spectrum}
\end{figure}

In the previous work \cite{JanhunenPOP2018}, linearly stable long wavelength components attributable to nonlinear processes were observed to arise in the 1D system. Energy cascade toward long wavelengths is also observed in 2D simulations, with the difference that in 2D the saturation and nonlinear stage is reached much more quickly due to the presence of the linearly unstable long wavelength MTSI mode. A salient feature observed in figure~\ref{fig:roots-sim} is that the MTSI becomes active only once the cyclotron modes have saturated, and the cyclotron modes respond to the saturation of the MTSI by resuming growth. This enhances the modulational nonlinear coupling through a faster linear response. The ion density $k_y$ spectrum at $r=1.35\,\text{cm}$, shown in figure~\ref{fig:ion-spectrum}, clearly demonstrates the progression towards lower-$k$ modes through inverse cascade, and emergence of the turbulent spectrum soon after the MTSI mode saturates.

\begin{figure}[htp]
\includegraphics[width=\columnwidth,clip]{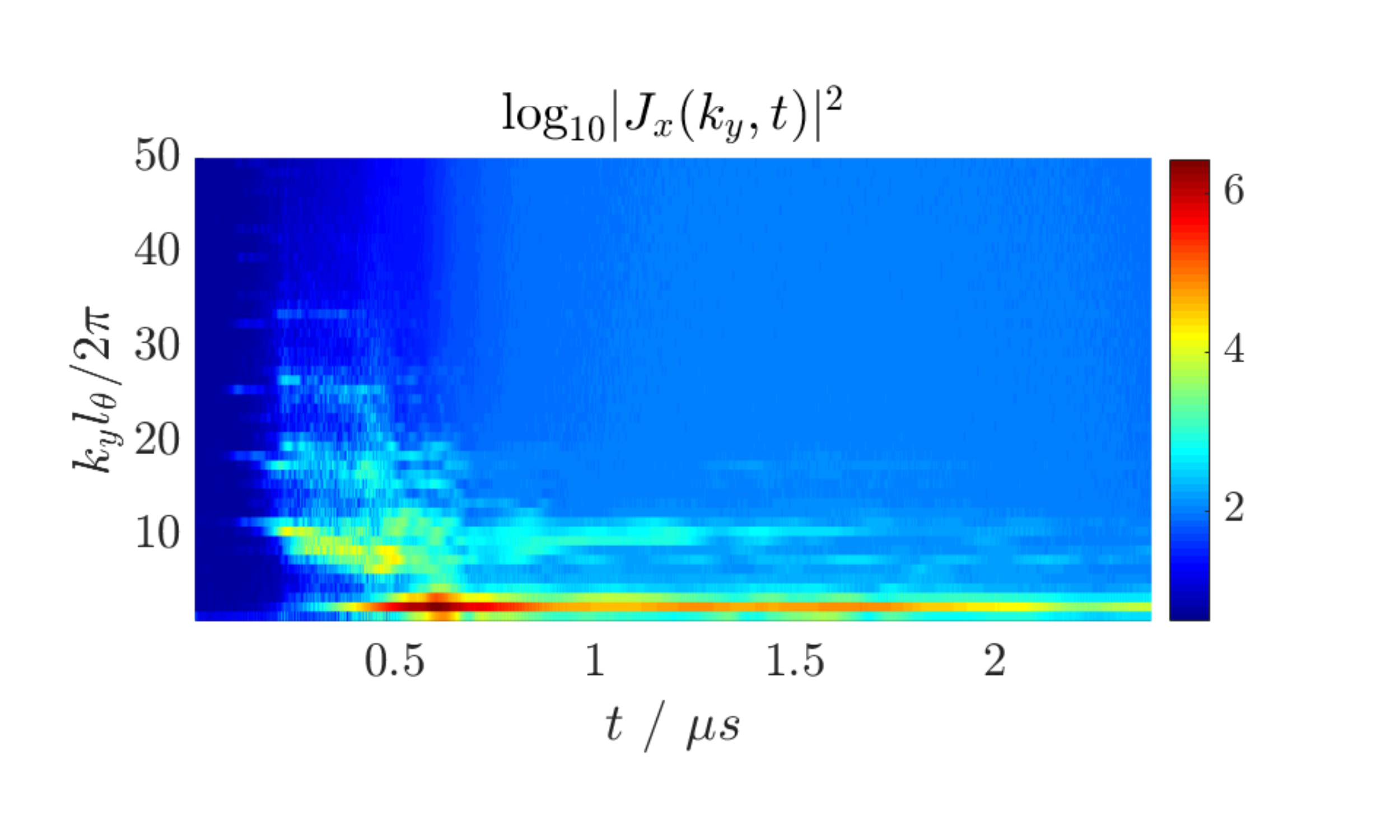}
\caption{Evolution of the anomalous current spectrum $\log_{10}{|\tilde{J}_z(k_y)|}$ over time at $z=L_r/4$ of the simulation. It is notable that the $n_y=1$ mode dominates soon after nonlinear saturation of the modes.}\label{fig:current-spectrum}
\end{figure}

Cascade to low-$k$ is even more drastic in electron density, where the $m>1$ resonances become all but absent. This process coincides with the generation of low-$k$ components in the anomalous current, and subsequent growth of net current. For the anomalous current, a similar spectrum may be obtained (figure~\ref{fig:current-spectrum}) illustrating  the initial cascade towards low-$k$ during the saturation of the ECDI modes. 

The present simulations encompass only a small fraction of the total Hall-effect thruster gap in the azimuthal direction, due to their computational expense, and even the longest wavelength modes remain short compared to the total gap circumference.

\section{Anomalous current}

 The axial anomalous current can be calculated from PIC2D as the relative charge flux between ions and electrons. The electric field exists only in the $(y,z)$-plane, but the axial anomalous current arises because of the cyclotron rotation in the $(x,y)$-plane, In particular, azimuthal electric field fluctuations (in $y$-direction) results in the electron displacement in $ x$-direction. The anomalous current $J_x$  obtained as the total charge flux in the $x$-direction is calculated directly from particles motion in the code.

\begin{figure}[htp]
\includegraphics[width=\columnwidth,clip]{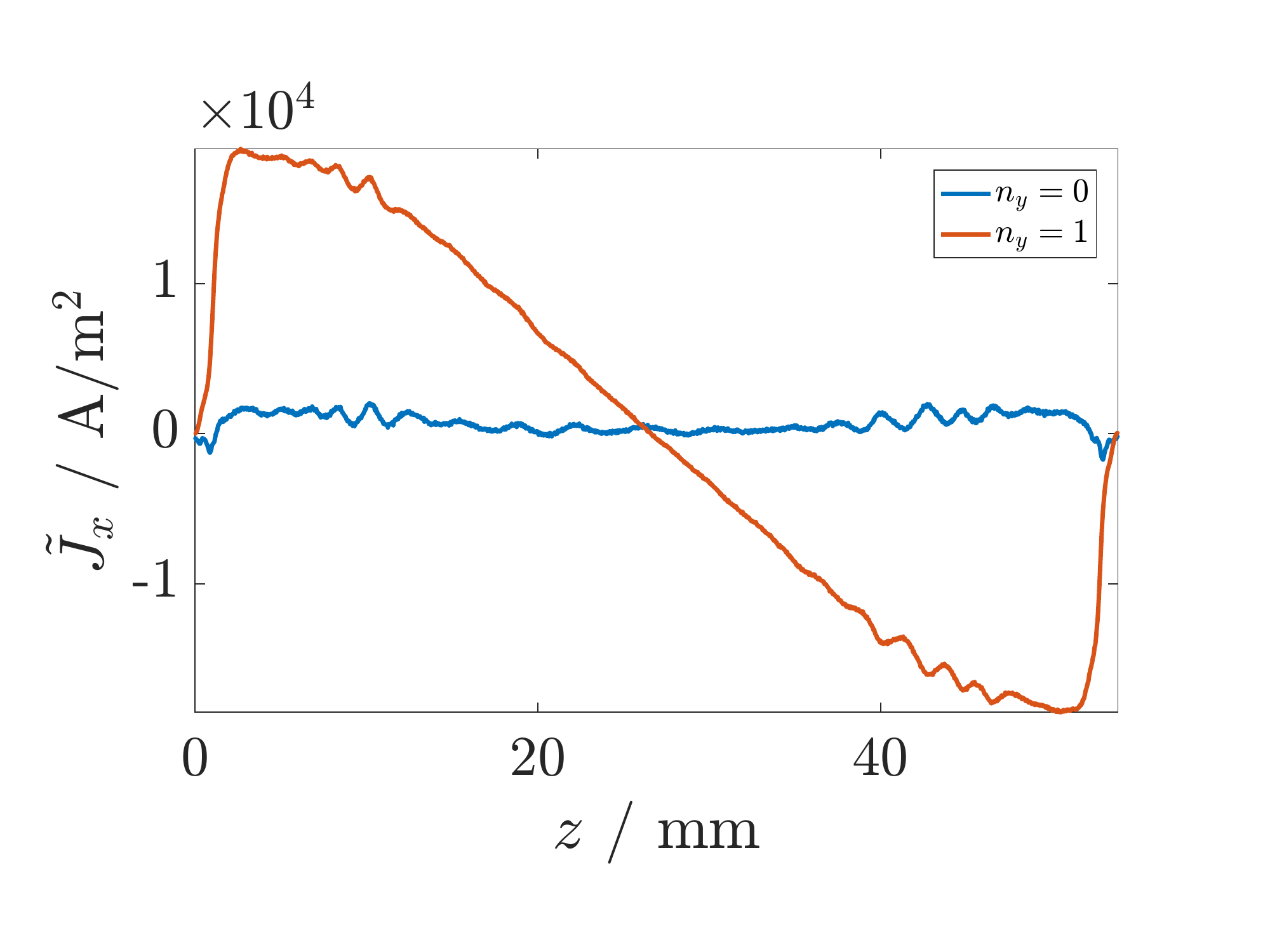}
\caption{Total anomalous current components $n_y=\{0,1\}$ for the nominal case. The total current is positive definite (blue curve), but the dominant $n_y=1$ component has a sheath-bounded half-wave radial profile corresponding to $n_z=1/2$.}\label{fig:current-mode-01}
\end{figure}

The inverse cascade is particularly apparent in the spectrum of the anomalous current (Fig.~\ref{fig:current-spectrum}). Another feature of the spectrum is the dominance of the $n_y=1$ mode that corresponds to the MTSI scale length. The mode also has the same radial envelope as the current profile (Fig.~\ref{fig:current-mode-01}).

\begin{figure}[htp]
\includegraphics[width=\columnwidth,clip]{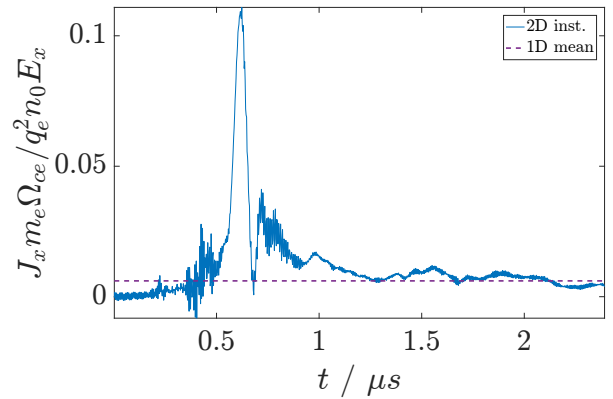}
\caption{Evolution of the total anomalous current density $J_z$ averaged over the simulation region. The anomalous current is normalized to the inverse of the Hall parameter\cite{LafleurPoP2016a}. The value obtained from 1D presented in Ref.\onlinecite{JanhunenPOP2018}, 1/164.}\label{fig:current-total}
\end{figure}

The spatial structure  of the anomalous current in 2D has some special features which are not possible in 1D simulations. The net axial current is associated with the $k_y=0$ component, which is plotted in figure~\ref{fig:current-total} as a function of time. Like in the 1D simulations, we observe that the anomalous current experiences an overshoot in the saturation stage, and settles down to a much lower level. The anomalous current spectrum is dominated by low-$k$ modes, particularly by the lowest mode available to the system in azimuthal direction, but also has a strong radial variation that is illustrated in figure~\ref{fig:current-mode-01}. Even though the MTSI creates a large transient in the total anomalous current, after the strong turbulent regime is established the net anomalous current falls to levels that are similar to those observed in 1D simulations. The long-wavelength features in parallel direction do not contribute to the total volume-averaged current, but could perhaps be observed in localized measurements as large alternating axial jets in the anomalous current.


\section{Sheath losses and decay of the plasma column}\label{sec:boundary}

The plasma in these simulations is bounded by dielectric boundaries in the $z$-direction. Charge accumulation on the dielectric surface was allowed and the displacement current in the dielectric with $\epsilon=4.5$ was taken into account using the model of Ref.~\onlinecite{SmolyakovPRL2013}.

\begin{figure}[htp]
  \includegraphics[width=\columnwidth,clip]{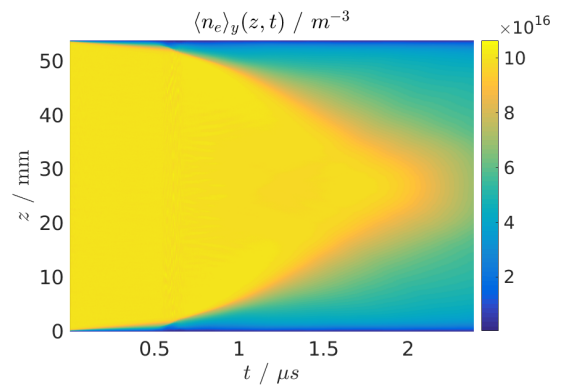}
\caption{Evolution of the azimuthal mean of the electron density as a function of time.}\label{fig:density}
\end{figure}

\begin{figure}[htp]
\includegraphics[width=\columnwidth,clip]{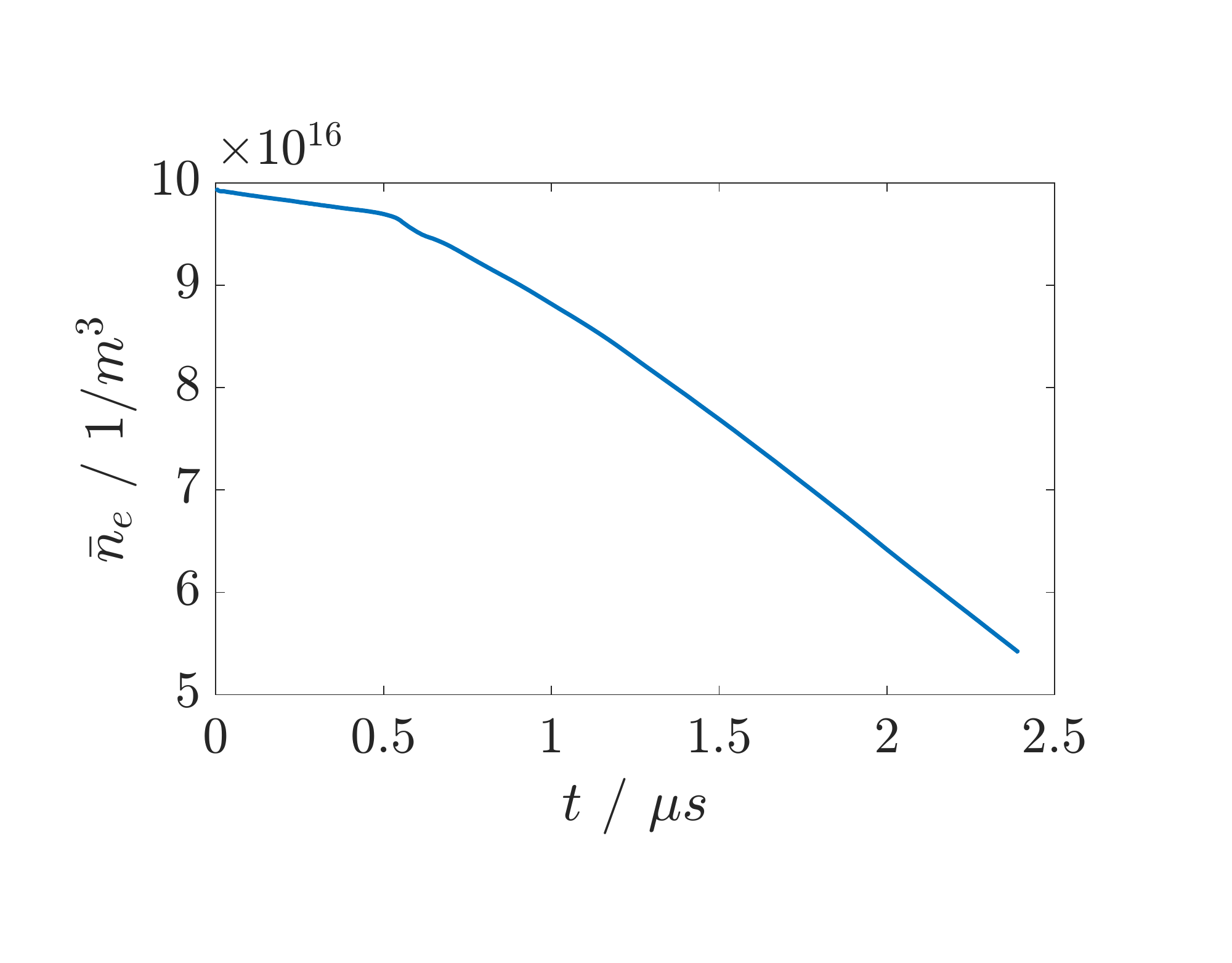}
\caption{Evolution of the volume-averaged electron density over time.}\label{fig:density-avg}
\end{figure}

A sheath (potential) develops very quickly in the simulation before any modes have had time to grow. After having formed, the sheath does not in fact expand much into the plasma (evident in Fig.~\ref{fig:density} ), but once the MTSI saturates we observe a faster decay rate of the plasma, likely due to the rapid electron heating during wave breaking. At $0.5\,\mu\text{s}$ the transition to a different regime is apparent particularly in figure~\ref{fig:density-avg}, where the rate of sheath losses increases drastically and the pre-sheath begins to rapidly expand into the plasma. This is the reason why our simulation box is larger than typical Hall-effect thruster gaps in the magnetic field direction --- without sources the expansion of the pre-sheath would happen too soon, and physics present in a steady-state thruster gap would be dynamically obscured.

Because the simulations presented in this paper do not have sources, it is an important question whether the decay due to sheath losses is significant enough to alter interpretation of the simulation results, and if so, in what manner. Decay of the density profile is shown in figures~\ref{fig:density} and \ref{fig:density-avg}, where the electron density profile in the magnetic field direction and the total volume averaged electron density are shown as a function of time. Decay of the plasma column and its effect on the modes is evident from Fig.~\ref{fig:stages}, where the modes terminate in $z$-direction where the density drop-off of the pre-sheath begins, except at later stages for a more gradual profile. Even at $2\,\mu\text{s}$ it appears that the nonlinear mode structure has remained largely intact (exhibiting characteristics of the cyclotron mode) although the plasma column has already decayed significantly (40\%). The profile is largely unaffected by pre-sheath expansion up to $1\,\mu\text{s}$ as seen from figure~\ref{fig:density}, and about 10\% of the electron density is lost so far.

Hence, we contend that the plasma column decay appears to have fairly little effect on the modes. Larger effects could be expected from the increase in electron temperature, which tends to modify the linear spectrum towards lower cyclotron resonances, as well as with secondary electrons (not included) that may even reverse the sheath. It is therefore likely that the robust features observed in these simulations would remain in the presence of sources and sinks too if a steady state can be achieved. There are some inherent difficulties in using sources (axial feeding and ionization) to achieve a steady-state even in 1D simulations, because of feedback between the modes and the sources (heating for ionization, transport for axial feeding).

Particularly in contrast to earlier work by H\'eron et al.\onlinecite{HeronPoP2013}, it is somewhat disappointing that the parallel heating mechanism has not been investigated to a greater degree in the literature. This makes it difficult to assign relative importance to the inclusion of secondary emission. 

\section{Summary and Conclusion}

  In earlier numerical studies of the electron cyclotron instability in 1D geometry Refs.~\onlinecite{LampePRL1971,LampePF1972a} it was found that the linear (exponential growth) stage of the fast beam cyclotron instability is saturated due to nonlinear turbulent broadening, which smears out the cyclotron resonances and the instability transitions into much slower ion-sound instability much like in the ordinary unmagnetized plasma. 
The authors of Refs.~\onlinecite{ForslundPRL1970,ForslundPF1972a} have performed analogous numerical studies for similar conditions and maintained that many properties of the observed instabilities are unlike those of the ion-sound mode of unmagnetized plasma. The apparent controversies from these simulations with regards to the importance and role of nonlinear electron diffusion, electron and ion trapping, as well as the role of the finite ion temperature have been discussed and contrasted at some length in two complementary papers, Refs.~\onlinecite{LampePF1972b} and \onlinecite{ForslundPF1972b}.   

The electron cyclotron drift instability driven by the electron current has been suggested as a possible candidate for enhanced electron transport in Hall truster in more recent Refs.~\onlinecite{AdamPoP2004,DucrocqPoP2006}. The authors of Refs.~\onlinecite{BoeufFP2014,BoeufJAP2017,BoeufIEPC2017,LafleurPoP2016a,LafleurPoP2016b,CroesPSST2017}  have performed a number of numerical simulations of ECDI and mostly concluded that the instability is analogous to the ion sound instability in absence of the magnetic field. Our recent 1D  simulations\cite{JanhunenPOP2018}, performed with higher resolution and with longer azimuthal simulation box have not confirmed this conclusion. In our simulations, we found that the criteria for the nonlinear resonance broadening and the destruction of cyclotron resonances is not satisfied and the instability proceeds as a coherent mode driven at the main  cyclotron resonance (of the reactive fluid type) $k_y v_0=\Omega_{ce}$ well into the nonlinear stage. Strong inverse energy cascade towards the longer wavelength was identified and it was shown that the anomalous current is dominated by the long wavelength modes.  

In 1D case, the transition to the ion sound regime 
may only occur due to the nonlinear resonance broadening and/or collisions. The role of numerical collisions was also discussed in Ref.~\onlinecite{ForslundPF1972b}. Numerical noise as well as particle re-injection (to limit the energy growth and mimic the axial direction) may also work as collisions. As it was discussed in the Section~\ref{sec:instability} above, in 2D geometry, when the direction along the magnetic field is resolved, the resonance thermal broadening due to a finite $k_z c_s$ may also facilitate the transition to the ion sound regime. However, arbitrary fluctuations are allowed at a sheath boundary, facilitating access to $k_z \approx 0$ regime where linear growth is largest.
In this paper, we have discussed the normal modes that are obtained from different limits of the full electrostatic dispersion relation, and compared them with the simulation results. Using the value of the effective $k_z$ obtained from simulations we have found that the linear dispersion relation predicts the discreet cyclotron resonance driven modes and the long wavelength MTSI modes which were also confirmed in the simulations. It appears therefore, that for our parameters full ion sound regime appears not to be realized, but instead a more effective regime of discrete cyclotron resonance instabilities and the long wavelength MTSI occur. The magnetic field remains to be a defining feature of these regimes (contrary to the unmagnetized ion sound regime) as also was concluded in Ref.~\onlinecite{ForslundPF1972b}.   

An important feature of the strong turbulence regime observed in our simulations is the difference in the ion and electron density fluctuations. While the electron density perturbations are rather benign and lower amplitude, the ion density perturbations have much larger amplitude and contain much larger short wavelength (non-quasineutral) content with $k_y \lambda_{De}\geq 1$. Such short wavelength modes do exhibit some interesting ion sound -like characteristics, such as the nonlinear $\omega_{pi}$ harmonics, nonlinear wave breaking, tendency for wave crests to become shock-like and the elastic-like collision of wave crests without interference. It is important to note that the intense fluctuations in the short wavelength part of the spectrum are less effective in supporting the anomalous electron current as well as for the electron heating. We note that rich short wavelength features are observed in ion density but not in electron density, which is much more smooth and coherent. This difference in the electron and ion response can be important for interpretation of the fluctuation diagnostics data\cite{TsikataPRL2015}. The coherent nature of the electron density wave and coherent electron-wave interaction is crucial for the electron heating mechanism \cite{ForslundPF1972b} that also precludes the application of the quasilinear theory\cite{MikhailenkopoP2003}. 

An interesting feature in our simulation is the strong $n_y=1$ component of the axial anomalous current, resulting in the alternating jets caused by the MTSI activity. The Modified Two-Stream Instability (MTSI) that was naturally absent from the 1D simulations is shown to amplify the inverse cascade tendency due to the long wavelength nature of the MTSI even in the linear regime. Development of the Modified Two-Stream Instability results in rapid parallel heating due to the finite electric field along the magnetic field. Intense parallel heating results in increased losses of high energy electrons into the sheath that serves as an additional saturation mechanism. Similar heating was also observed in Ref.~\cite{HeronPoP2013}. Their simulations also suggest that effects of secondary emission could significantly increase anomalous transport, but unfortunately restrict wave vector space for heating studies. They too observe modulations in the sheath, however.
In our simulations, the anomalous electron transport sets at the level similar to that in our 1D simulations\cite{JanhunenPOP2018}, perhaps due to the absence of secondary emission.    

\section{Supplementary Material}

We provide movies of three quantities discussed in the paper: $\delta{n}_i$, $\delta{n}_e$ and $T_e$ over the simulation box $L_r\times l_\theta$. They illustrate the dynamics observed and reported in this paper.

\begin{acknowledgements}
This work was supported in part by NSERC Canada and the Air Force Office of Scientific Research under awards number FA9550-18-1-0132 and FA9550-15-1-0226. Computational resources from ComputeCanada/WestGrid were used in this work.
\end{acknowledgements}

\appendix
\section{On linear growth rate analysis and other techniques}\label{app:determine}

It is possible to determine all the linear growth rates directly from the spectrogram obtained from the simulation, when the simulations are run with good resolution both in space and particle number to have the modes be well resolved with $k_{\Delta x} \lambda_{De}/4 \geq 1$ and deep in the linear regime. Growth rates from the non-linear 2D simulation are obtained by discrete Fourier decomposition in space, and expressing the Fourier coefficients as a function of time:
\begin{equation}
\phi(z,y; t)=\frac{1}{N}\sum_{k=-N/2}^{N/2} c_k(z,t) \exp{i k y},
\end{equation}
and with the usual expression for a traveling wave, we have $c_{k}(z,t)=\tilde{c}_{k}(z)\exp{i\int\omega(t)\,dt)}$, where $\omega(t)$ is complex-valued. Particularly in the linear regime where $\omega=\omega_r+i\gamma$ we may obtain the growth rate and frequency as a linear fit to the main branch of the complex logarithm of the Fourier coefficients as $i \omega(t)=d/dt\log(c_k)$.

Waves are largely observed to propagate in the periodic $y$-direction, so fluctuations of quantities are measured against the $y$-mean for convenience. Namely, the subtracted ``average'', here for electron density, is defined by $\langle n_e\rangle_y(z,t)=\frac{1}{l_\theta}\int_0^{l_\theta} n_e(y,z,t) \,dy$. Using this density as the reference also suppresses the large drop that is observed in the sheath.

In figure~\ref{fig:roots-sim} we illustrate this process as performed on our simulation data. The MTSI mode (first peak) is well represented though, showing the importance of good statistics. To emphasize this point, we ran a case for $T_e\approx 0$ with the 2D code, and a case with 1D code with very good statistics (Fig.~\ref{fig:sim-gamma}) that gives a very good value for both the growth rate and frequency of the Buneman instability (equation~\ref{mbtsi}). In the 1D simulation we had $N_p/N_g=40000$ particles per cell and $N_g=3400$ cells, allowing for resolution that is difficult to get with 2D simulations due to computational limitations. It would be possible to extend the growth region for ECDI by decreasing the noise level by increasing the particle number, but the simulations up to the nonlinear stage would become unfeasible. The 2D simulations use electron sub-cycling with a modest ratio of $N_{cyc}=3$ electron steps per ion step.

\begin{table}[htp]
\begin{tabular}{|c|c|}\hline
Symbol & Value\\\hline
$T_e$ & 10 eV\\
$n_{e,i}$ & $10^{17}$\\
$\lambda_{De}/\{\Delta{y},\Delta{z}\}$& $4/\sqrt{2}$\\
$N_p/N_g$ &800\\
$l_\theta$/$\Delta{y}$&512\\
$L_r/\Delta{z}$&2048\\
$B_0$&0.02 T\\
$E_0$&20 kV/m\\
$N_{cyc}$&3\\\hline
\end{tabular}
\caption{Numerical parameters for the 2D ECDI simulations.}\label{table:numerical_params}
\end{table}

\bibliography{MyReferences}

\end{document}